\documentclass[onecolumn,superscriptaddress,showpacs,nofootinbib]{revtex4}
 \usepackage{graphicx}

\usepackage{bm,color}

\begin{document}

\title{Linear response theory for hydrodynamic and kinetic equations with long-range interactions}

% First author block:
\author{Pierre-Henri Chavanis}
%\email{chavanis@irsamc.ups-tlse.fr}
% \homepage{An author's web page; optional}
\affiliation{ Laboratoire de Physique Th\'eorique (IRSAMC), CNRS and UPS,  Universit\'e de Toulouse, F-31062 Toulouse, France}
% You may list several affiliation, using separate commands for each:
%\affiliation{The third affiliation is shared by both co-authors}

% For other authors please repeat the author block as needed
%\author{Second Author}
% Note how REVTeX 4 deals with identical affiliations
%\affiliation{The third affiliation is shared by both co-authors}

\begin{abstract}
We apply the linear response theory to systems with long-range
interactions described by hydrodynamic equations such as the Euler,
Smoluchowski, and damped Euler equations. We analytically determine
the response of the system submitted to a pulse and to a step
function. We compare these results with those obtained for
collisionless systems described by the Vlasov equation.  We show that,
in the linear regime, the evolution of a collisionless system (Vlasov)
with the waterbag distribution is the same as the evolution of a
collision-dominated gas without dissipation (Euler). In this analogy,
the maximum velocity of the waterbag distribution plays the role of
the velocity of sound in the corresponding barotropic gas. When
submitted to a step function, these systems exhibit permanent
oscillations. Other distributions exhibit Landau damping and relax
towards a steady state. We illustrate this behaviour with the Cauchy
distribution which can be studied analytically. We apply our results
to the HMF model and obtain a generalized Curie-Weiss law for the
magnetic susceptibility. Finally, we compare the linear response
theory to the initial value problem for the linearized Vlasov equation
and report a case of algebraic damping of the initial perturbation.

\end{abstract}

% Insert suggested PACS numbers (up to 4) in braces.
% The PACS (Physics and Astronomy Classification Scheme)
% can be accessed on the web at http://www.aip.org/pacs/
%\pacs{\textcolor{red}{47.10.A-,47.15.ki}}

% Insert keywords (up to about 4) in braces; optional.
% \keywords{Up to four keywords}

\maketitle

% Here the text of your article begins

\section{\label{intro}Introduction}
% References should be done using the \cite, \ref, and \label commands.
% Put \label in argument of \section for cross-referencing like this:
%\section{\label{}}

The linear response theory is a simple and powerful tool for studying the stability of a dynamical system and its response to an external perturbation. Kubo \cite{kubo} used it to determine the variation of the average value of an ``observable'' due to an applied ``force''. The linear response theory found a lot of applications in statistical mechanics and  kinetic theory \cite{kubobook}, and in the theory of simple liquids \cite{hansen}. Its most important applications concern the derivation of expressions for the transport coefficients of hydrodynamics, including the electrical and thermal conductivity, and the mobility of a Brownian particle immersed in a fluid.

The linear response theory has also been applied to long-range interacting systems \cite{balescubook,houches,bt,cdr}. In particular, it has been used to compute the friction force experienced by a star passing through a globular cluster \cite{marochnik,kalnajs,kandrup} or the drift of a point vortex moving in a background shear \cite{drift,sd}. These results can be generalized  to other systems with long-range interactions \cite{hb4}. Recently, the linear response theory has been applied to the Hamiltonian Mean Field (HMF) model \cite{ar} in order to determine how a long-range system in a quasistationary state (QSS) responds to an external perturbation \cite{patelli,oyresponse}. In the examples discussed above, the linear response theory is based on the Liouville equation for a Hamiltonian system of particles in interaction, or on the Vlasov equation for a collisionless gas. In the present paper, we apply the linear response theory to a collision-dominated gas described by hydrodynamic equations such as the Euler, Smoluchowski, and damped Euler equations. Specifically, we consider the response of this system to a weak external perturbation, and we compare the results with those obtained for a collisionless gas.

The paper is organized as follows. In Section \ref{sec_pr}, we recall general relations satisfied by the polarization and response functions. As an application, we consider the response of the system to a pulse and to a step function. In Section \ref{sec_de}, we apply the linear response theory to a collision-dominated gas described by hydrodynamic equations. For the sake of generality, we consider the damped Euler equation including a friction force $-\xi {\bf u}$ proportional to the velocity. For $\xi=0$, we recover the pure Euler equation describing an ideal gas, and for $\xi\rightarrow +\infty$, we get the Smoluchowski equation describing the overdamped motion of Brownian particles. Explicit expressions of the response and polarization functions are obtained for these systems. In Section \ref{sec_vlasov}, we compare these results with those obtained for a collisionless gas described by the  Vlasov equation. We show that, in the linear regime,  the evolution of a collisionless system (Vlasov) with the waterbag distribution is the same as the evolution of a collision-dominated gas without dissipation (Euler). In this analogy, the maximum velocity of the waterbag distribution plays the role of the velocity of sound in the corresponding barotropic  gas. When submitted to a step function, these systems exhibit permanent oscillations. In Section \ref{sec_cauchy}, we treat the case of the Cauchy distribution for which analytical results can be obtained.  When submitted to a step function, this distribution exhibits Landau damping and relaxes towards a steady state. In Section \ref{sec_hydro}, we study the evolution of the perturbed density in the linear regime and establish hydrodynamic equations for a collision-dominated gas and a collisionless system. In Section \ref{sec_asy}, we consider the asymptotic distribution of a stable system submitted to a step function and show that it coincides with the steady distribution of this system under a weak external field.  In Section \ref{sec_app}, we apply our general results to the HMF model and consider specific distribution functions such as the isothermal, polytropic, waterbag, and Fermi-Dirac distributions. We show that the magnetic susceptibility of the equilibrium state is given by a generalized Curie-Weiss law. In Section \ref{sec_iv}, we compare the response of a collisionless system submitted to a weak external potential with the evolution of an isolated system submitted to an initial disturbance (the so-called initial value problem of the linearized Vlasov equation) and we report a case of algebraic damping. In Appendix \ref{sec_corr}, we discuss the connection between a collisionless system described by a distribution function depending only on the individual energy and the corresponding barotropic gas. In Appendix \ref{sec_pgh}, we illustrate some results for Coulombian plasmas, self-gravitating systems, and for the HMF model.

\section{Polarization and response functions}
\label{sec_pr}

\subsection{General results}

We examine the response of a spatially homogeneous system at equilibrium to a small external potential $\Phi_e({\bf r},t)$. We follow the presentation given by Binney and Tremaine \cite{bt}. The perturbation caused by the external potential may be written as
\begin{equation}
\label{pr1}
\delta\Phi({\bf r},t)=\int d{\bf r}' dt'\, R({\bf r}-{\bf r}',t-t') \Phi_e({\bf r}',t'),
\end{equation}
where $R({\bf x},\tau)$ is the response function. Causality requires
that $R({\bf x},\tau)=0$ for $\tau<0$. We introduce the polarization
function $P({\bf x},\tau)$ which relates the perturbation to the total
potential:
\begin{equation}
\label{pr2}
\delta\Phi({\bf r},t)=\int d{\bf r}' dt'\, P({\bf r}-{\bf r}',t-t') \left\lbrack \Phi_e({\bf r}',t')+\delta \Phi({\bf r}',t')\right\rbrack.
\end{equation}
Once again, $P({\bf x},\tau)=0$ for $\tau<0$. The polarization function corresponds to the response of the system when collective effects are neglected. Indeed, if we neglect the self-interaction of the system  $\delta\Phi$ in the calculation of the response function $R$ (see below), we obtain the polarization function $P$.

Since the external potential is introduced at $t=0$ (say), it is convenient to use Laplace transforms in time and Fourier transforms in space. The Fourier-Laplace transform of the perturbed potential  $\delta \Phi({\bf r},t)$ is defined by
\begin{equation}
\delta \tilde \Phi({\bf k},\omega)=\int \frac{d{\bf r}}{(2\pi)^d}\int_{0}^{+\infty}dt\, e^{-i({\bf k}\cdot{\bf r}-\omega t)}\delta \Phi({\bf r},t).
\label{pr3}
\end{equation}
This expression for the Laplace transform is valid for ${\rm Im}(\omega)$ sufficiently large. For the remaining part of the complex $\omega$ plane, it is defined by an analytic continuation. The inverse Fourier-Laplace transform is
\begin{equation}
\delta \Phi({\bf r},t)=\int d{\bf k}\int_{\cal C}\frac{d\omega}{2\pi}\, e^{i({\bf k}\cdot{\bf r}-\omega t)}\delta\tilde \Phi({\bf k},\omega),
\label{pr4}
\end{equation}
where the Laplace contour ${\cal C}$ in the complex $\omega$ plane must pass above all poles of the integrand. Taking the Fourier transforms of Eqs.  (\ref{pr1}) and (\ref{pr2}), we obtain
\begin{equation}
\label{pr5}
\delta\hat{\Phi}({\bf k},t)=\int_{0}^{t} dt'\, R({\bf k},t-t')\hat{\Phi}_{e}({\bf k},t'),
\end{equation}
\begin{equation}
\label{pr6}
\delta\hat{\Phi}({\bf k},t)=\int_{0}^{t} dt'\, P({\bf k},t-t')\left\lbrack \hat{\Phi}_{e}({\bf k},t')+\delta\hat{\Phi}({\bf k},t')\right\rbrack.
\end{equation}
Taking the Fourier-Laplace transform of Eqs. (\ref{pr1}) and (\ref{pr2}), we get
\begin{equation}
\label{pr7}
\delta\tilde\Phi({\bf k},\omega)=R({\bf k},\omega) \tilde\Phi_e({\bf k},\omega),
\end{equation}
\begin{equation}
\label{pr8}
\delta\tilde\Phi({\bf k},\omega)=P({\bf k},\omega)\left\lbrack \tilde\Phi_e({\bf k},\omega)+\delta\tilde\Phi({\bf k},\omega)\right\rbrack.
\end{equation}
The Fourier-Laplace transforms of the response and polarization functions are related to each other by
\begin{equation}
\label{pr9}
R({\bf k},\omega)=\frac{P({\bf k},\omega)}{1-P({\bf k},\omega)},\qquad P({\bf k},\omega)=\frac{R({\bf k},\omega)}{1+R({\bf k},\omega)}.
\end{equation}
If we define the dielectric function by
\begin{equation}
\label{pr10}
\epsilon({\bf k},\omega)=1-P({\bf k},\omega),
\end{equation}
we obtain
\begin{equation}
\label{pr11}
R({\bf k},\omega)=\frac{1-\epsilon({\bf k},\omega)}{\epsilon({\bf k},\omega)}.
\end{equation}

\subsection{The response to a pulse}
\label{sec_pulss}

We consider the response of the system to a ``pulse'' localized at $t=0$. It can be represented by a Dirac distribution
\begin{equation}
\label{pr12}
\hat{\Phi}_e({\bf k},t)=\delta(t).
\end{equation}
We have directly written the Fourier transform of the external potential and, for simplicity, we have assumed that the perturbation is independent on the mode ${\bf k}$ (this situation can be straightforwardly generalized by multiplying the results by some amplitude $\hat{\Phi}_e({\bf k})$). The Laplace transform of the external potential is
\begin{equation}
\label{pr13}
\tilde{\Phi}_e({\bf k},\omega)=1.
\end{equation}
According to Eqs. (\ref{pr7}) and (\ref{pr13}), the perturbation caused by a pulse is equal to the response function:
\begin{equation}
\label{pr14}
\delta\tilde{\Phi}({\bf k},\omega)=R({\bf k},\omega).
\end{equation}
Taking the inverse Laplace transform of this expression, and using Eq. (\ref{pr11}), we obtain
\begin{equation}
\label{pr15}
\delta\hat{\Phi}({\bf k},t)=\int_{\cal C}\frac{d\omega}{2\pi}\, e^{-i\omega t}\frac{1-\epsilon({\bf k},\omega)}{\epsilon({\bf k},\omega)}.
\end{equation}
The poles of the integrand correspond to the complex pulsations $\omega_{\alpha}({\bf k})$ for which the dielectric function vanishes: $\epsilon({\bf k},\omega_{\alpha}({\bf k}))=0$. This defines the dispersion relation. The evolution of the perturbation depends on the position of the zeros of the dielectric function in the complex plane. Using the Cauchy residue theorem, we have
\begin{equation}
\label{pr16}
\delta\hat{\Phi}({\bf k},t)=-i\sum_{\alpha} e^{-i\omega_{\alpha}({\bf k}) t} \left\lbrack {\rm Res} \frac{1-\epsilon({\bf k},\omega)}{\epsilon({\bf k},\omega)}\right \rbrack_{\omega=\omega_{\alpha}({\bf k})},
\end{equation}
where the sum runs over the whole set of poles and we have assumed, for simplicity, that the singularities are simple poles. In the following, we shall omit the subscript $\alpha$ for brevity. If at least one zero $\omega$ of the dielectric function lies in the upper half plane (i.e. $\omega_i>0$), the system is unstable, and the perturbation grows exponentially rapidly with the rate $(\omega_i)_{max}$ corresponding to the zero with the largest value of the imaginary pulsation. If all the zeros $\omega$ of the dielectric function strictly lie in the lower half-plane (i.e. $\omega_i<0$), the system is stable, and the perturbation decays to zero exponentially rapidly with the rate $|\omega_i|_{min}$ corresponding to the zero with the smallest value of the imaginary pulsation in absolute value.  Finally, if the zeros lie on the real axis (i.e. $\omega_i=0$), the system is marginally stable and the perturbation displays an oscillating behavior with the pulsation $\omega_r$. For more details, we refer to \cite{balescubook,bt}.

\subsection{The response to a step function}

We consider the response of the system to a constant potential that is ``switched on'' suddenly at $t=0$. It can be represented by a step function
\begin{equation}
\label{pr17}
\hat{\Phi}_e({\bf k},t)=H(t)\hat{\Phi}_e({\bf k}),
\end{equation}
where $H(t)=0$ for $t<0$ and $H(t)=1$ for $t>0$ (Heaviside function). The Laplace transform of the external potential is
\begin{equation}
\label{pr18}
\tilde{\Phi}_e({\bf k},\omega)=\frac{i}{\omega}\hat{\Phi}_e({\bf k}).
\end{equation}
According to Eqs. (\ref{pr7}) and (\ref{pr18}), the perturbation caused by a step function is
\begin{equation}
\label{pr19}
\delta\tilde{\Phi}({\bf k},\omega)= R({\bf k},\omega)\frac{i}{\omega}\hat{\Phi}_e({\bf k}).
\end{equation}
Taking the inverse Laplace transform of this expression, and using Eq. (\ref{pr11}), we obtain
\begin{equation}
\label{pr20}
\delta\hat{\Phi}({\bf k},t)=\hat{\Phi}_e({\bf k})\int_{\cal C}\frac{d\omega}{2\pi}\, e^{-i\omega t}\frac{1-\epsilon({\bf k},\omega)}{\epsilon({\bf k},\omega)}\frac{i}{\omega}.
\end{equation}
The integrand presents a pole at $\omega=0$ that gives rise to a constant term $(1-\epsilon({\bf k},0))/\epsilon({\bf k},0)$. On the other hand, the temporal evolution of the perturbation depends on the position of the zeros of the dielectric function $\epsilon({\bf k},\omega)$ in the complex plane. Using the Cauchy residue theorem, we have
\begin{equation}
\label{pr21}
\frac{\delta\hat{\Phi}({\bf k},t)}{\hat{\Phi}_e({\bf k})}=\frac{1-\epsilon({\bf k},0)}{\epsilon({\bf k},0)}+\sum_{\alpha} e^{-i\omega_{\alpha}({\bf k}) t} \left\lbrack {\rm Res} \frac{1-\epsilon({\bf k},\omega)}{\epsilon({\bf k},\omega)\omega}\right \rbrack_{\omega=\omega_{\alpha}({\bf k})},
\end{equation}
where we have assumed, for simplicity, that the singularities are simple poles. If at least one zero $\omega$ of the dielectric function lies in the upper half plane (i.e. $\omega_i>0$), the system is unstable, and the perturbation grows exponentially rapidly with the rate $(\omega_i)_{max}$. If all the zeros $\omega$ of the dielectric function $\epsilon({\bf k},\omega)$ strictly lie in the lower half-plane (i.e. $\omega_i<0$), the late time evolution of the perturbation $\delta\hat{\Phi}({\bf k},t)$ is dominated by the pole at $\omega=0$. As a result, the perturbation tends to the asymptotic value
\begin{equation}
\label{pr22}
\delta\hat{\Phi}_{\infty}({\bf k})=\frac{1-\epsilon({\bf k},0)}{\epsilon({\bf k},0)}\hat{\Phi}_e({\bf k}),
\end{equation}
for $t\rightarrow +\infty$.  Finally, if the zeros lie on the real axis (i.e. $\omega_i=0$), the perturbation displays an oscillating behavior about the value (\ref{pr22}) with the pulsation $\omega_r$.

\section{The damped barotropic Euler equation }
\label{sec_de}

We consider a system of particles with long-range interactions interacting via a binary potential $u(|{\bf r}-{\bf r}'|)$. We assume that the particles also experience collisions due to short-range interactions. In the absence of dissipation, this system is described by the  Euler equations. These hydrodynamic equations are the correct description of a collision-dominated gas. For the sake of generality, we also allow for the possibility that the particles of the system move in an inert medium and experience a friction force proportional to their velocity. This is the case, for example, for colloidal particles immersed in a fluid. In order to study these two situations in a unified manner, we consider the damped Euler equations \cite{nfp}:
\begin{equation}
\label{de1}
\frac{\partial \rho}{\partial t}+\nabla\cdot (\rho {\bf u})=0,
\end{equation}
\begin{equation}
\label{de2}
\rho\left\lbrack \frac{\partial {\bf u}}{\partial t}+({\bf u}\cdot \nabla){\bf u}\right \rbrack=-\nabla p-\rho\nabla\Phi-\rho\nabla\Phi_e-\xi\rho{\bf u},
\end{equation}
\begin{equation}
\label{de3}
\Phi({\bf r},t)=\int u(|{\bf r}-{\bf r}'|)\rho({\bf r}',t)\, d{\bf r}',
\end{equation}
where $\xi$ is the friction coefficient. We have distinguished the potential $\Phi({\bf r},t)$ produced by the particles from the external potential $\Phi_e({\bf r},t)$. To close the system of equations, we need to specify the equation of state. We consider a barotropic gas in which the pressure is a function of the density: $p({\bf r},t)=p\lbrack \rho({\bf r},t)\rbrack$. The specification of the equation of state $p=p(\rho)$ completely closes the system of equations.

The friction coefficient $\xi$ measures the strength of  dissipative effects. For $\xi=0$, Eqs. (\ref{de1})-(\ref{de3}) reduce to the Euler equations for a perfect fluid. Alternatively, in the strong friction limit $\xi\rightarrow +\infty$, we can neglect inertial effects in Eq. (\ref{de2}) and obtain
\begin{equation}
\label{de4}
\rho{\bf u}\simeq -\frac{1}{\xi}(\nabla p+\rho\nabla\Phi+\rho\nabla\Phi_e).
\end{equation}
Substituting this relation in the equation of continuity (\ref{de1}),  we get the generalized Smoluchowski equation \cite{nfp}:
\begin{equation}
\label{de5}
\frac{\partial \rho}{\partial t}=\nabla\cdot \left\lbrack \frac{1}{\xi}\left (\nabla p+\rho\nabla\Phi+\rho\nabla\Phi_e\right )\right\rbrack,
\end{equation}
\begin{equation}
\label{de6}
\Phi({\bf r},t)=\int u(|{\bf r}-{\bf r}'|)\rho({\bf r}',t)\, d{\bf r}'.
\end{equation}
This equation describe the dynamics of Langevin particles in interaction in a strong friction limit. Therefore, the damped Euler equation makes the link between the Euler equation ($\xi=0$)  and the Smoluchowski equation ($\xi\rightarrow +\infty$). For a more detailed discussion of these equations, we refer to \cite{longshort}.

When the potential of interaction $u(|{\bf r}-{\bf r}'|)$ is the gravitational potential, the mean field equation (\ref{de3}) or (\ref{de6}) reduces to the Poisson equation. The Euler-Poisson system describes a self-gravitating collision-dominated gas like a barotropic star for example \cite{bt}. The damped Euler-Poisson system  may describe the dynamics of dust particles in the solar nebula \cite{aa}. In the strong friction limit, the generalized Smoluchowski-Poisson system describes self-gravitating Langevin particles \cite{cslangevin}. Finally, the case of a cosine potential of interaction in 1D has been considered in relation to the HMF \cite{ar} and BMF (p. 86 of \cite{cvb}) models.

\subsection{The linearized damped Euler equations}
\label{sec_lde}

We consider a system in a steady state with uniform density $\rho$ and examine its response to a small external potential $\Phi_e({\bf r},t)$. Since the perturbation is small, we can develop a linear response theory. The linearized damped Euler equations are
\begin{equation}
\label{lde1}
\frac{\partial \delta\rho}{\partial t}+\rho\nabla\cdot {\bf u}=0,
\end{equation}
\begin{equation}
\label{lde2}
\rho \frac{\partial {\bf u}}{\partial t}=-c_s^2\nabla \delta\rho-\rho\nabla\delta\Phi-\rho\nabla\Phi_e-\xi\rho{\bf u},
\end{equation}
\begin{equation}
\label{lde3}
\delta\Phi({\bf r},t)=\int u(|{\bf r}-{\bf r}'|)\delta\rho({\bf r}',t)\, d{\bf r}',
\end{equation}
where $c_s^2=p'(\rho)$ is the velocity of sound in the gaseous medium. Taking the Fourier-Laplace transform of these equations, we get
\begin{equation}
\label{lde4}
-i\omega\delta\tilde\rho+i\rho {\bf k}\cdot  \tilde{\bf u}=0,
\end{equation}
\begin{equation}
\label{lde5}
-i\omega \rho \tilde{\bf u}=-c_s^2 i\delta\tilde\rho{\bf k}-i\rho\delta\tilde\Phi{\bf k}-i\rho\tilde\Phi_e{\bf k}-\xi\rho\tilde{\bf u},
\end{equation}
\begin{equation}
\label{lde6}
\delta\tilde\Phi=(2\pi)^d \hat{u}(k)\delta\tilde\rho.
\end{equation}
Taking the scalar product of Eq. (\ref{lde5}) with ${\bf k}$, and using Eqs. (\ref{lde4}) and (\ref{lde6}), we find that
\begin{equation}
\label{pr14mar}
\delta\tilde{\Phi}({\bf k},\omega)=R({k},\omega)\tilde\Phi_e({\bf k},\omega),
\end{equation}
with the response function
\begin{equation}
\label{lde7}
R({k},\omega)=\frac{(2\pi)^d\hat{u}(k) k^2 \rho}{\omega^2+i\xi \omega-c_s^2k^2-(2\pi)^d\hat{u}(k)k^2\rho}.
\end{equation}
Using Eqs. (\ref{pr9}) and (\ref{pr10}), we obtain the polarization function
\begin{equation}
\label{lde8}
P({k},\omega)=\frac{(2\pi)^d\hat{u}(k) k^2 \rho}{\omega^2+i\xi \omega-c_s^2k^2},
\end{equation}
and the dielectric function
\begin{equation}
\label{lde9}
\epsilon({k},\omega)=1-\frac{(2\pi)^d\hat{u}(k) k^2 \rho}{\omega^2+i\xi \omega-c_s^2k^2}.
\end{equation}

\subsection{The solution of the dispersion relation}
\label{sec_dr}

The dispersion relation $\epsilon({k},\omega)=0$ can be written explicitly as
\begin{equation}
\label{dr1}
\omega^2+i\xi\omega-\omega_0^2(k)=0,
\end{equation}
where $\omega_0^2(k)=c_s^2k^2+(2\pi)^d\hat{u}(k)k^2\rho$. The dispersion relation determines the complex pulsation $\omega$ as a function of the wavenumber ${k}$. We find that
\begin{equation}
\label{dr3}
\omega=\frac{-i\xi\pm\sqrt{\Delta(k)}}{2},
\end{equation}
where $\Delta(k)=4\omega_0^2(k)-\xi^2$. If $\omega_0^2(k)>0$, the imaginary part of the complex pulsation is negative, implying stability. If $\omega_0^2(k)<0$, the imaginary part of the complex pulsation with the sign $+$  is positive, implying instability. Therefore, the  system is stable with respect to a perturbation with wavenumber $k$ when
\begin{equation}
\label{dr5}
\epsilon(k,0)=1+\frac{(2\pi)^d\hat{u}(k)\rho}{c_s^2} > 0,
\end{equation}
and unstable otherwise. This stability criterion can also be obtained from the Nyquist theorem \cite{nyquist}.  It provides a generalization of the Jeans instability criterion in astrophysics \cite{bt}. For repulsive potentials $\hat{u}(k)>0$ (as in plasma physics), the system is always stable. For attractive potentials $\hat{u}(k)<0$ (as in astrophysics), the system may be unstable to some wavelengths.

A mode is a perturbation that can be sustained without external
forces. It is therefore the solution of Eqs. (\ref{lde1})-(\ref{lde3})
with $\Phi_e=0$. The modes are usually of the form $\delta\rho\propto
e^{i({\bf k}\cdot {\bf r}-\omega(k) t)}$ where $\omega(k)$ is the
solution of the dispersion relation (\ref{dr1}). When $\Delta(k)\neq
0$, $\delta\rho$ is a linear combination of the two modes
$\omega_{\pm}(k)$ given by Eq. (\ref{dr3}). When $\Delta(k)=0$,
$\omega_{+}(k)=\omega_{-}(k)=-i\xi/2$ and the solution of
Eqs. (\ref{lde1})-(\ref{lde3}) is a linear combination of
$\delta\rho\propto e^{-\xi t/2} e^{i{\bf k}\cdot {\bf r}}$ and
$\delta\rho\propto t e^{-\xi t/2} e^{i{\bf k}\cdot {\bf r}}$ (this may
be seen, for example, on the hydrodynamic equation (\ref{h3}) that is
equivalent to Eqs. (\ref{lde1})-(\ref{lde3})).

For the Euler equation ($\xi=0$), the dispersion relation reduces to $\omega^2=\omega_0^2(k)$. In the stable case, the perturbation oscillates with a pulsation $\omega_r=\pm\omega_0(k)$. In the unstable case, the perturbation grows exponentially rapidly with a growth rate $\omega_i=\sqrt{-\omega_0^2(k)}>0$ (the second mode is damped exponentially rapidly with a damping rate $\omega_i=-\sqrt{-\omega_0^2(k)}<0$).

For the Smoluchowski equation ($\xi\rightarrow +\infty$), the dispersion relation reduces to $i\xi\omega=\omega_0^2(k)$. In the stable case, the perturbation is damped exponentially rapidly with a damping rate  $\omega_i=-\omega_0^2(k)/\xi<0$. In the unstable case, the perturbation grows exponentially rapidly with a growth rate $\omega_i=-\omega_0^2(k)/\xi>0$.

We now consider the damped Euler equation. In the stable case, we have to distinguish three cases: If $\Delta(k)>0$, the perturbation oscillates with a pulsation $\omega_r=\pm\sqrt{\Delta(k)}/2$ while being  damped exponentially rapidly with a damping rate $\omega_i=-\xi/2<0$; if $\Delta(k)<0$, the perturbation is damped exponentially rapidly with a damping rate $\omega_i=(-\xi+\sqrt{-\Delta(k)})/2<0$ (the second mode is damped more rapidly at a rate $\omega_i=(-\xi-\sqrt{-\Delta(k)})/2<0$); if $\Delta(k)=0$, the temporal evolution of the perturbation behaves as $te^{-\xi t/2}$. In the unstable case, the perturbation grows exponentially rapidly with a growth rate $\omega_i=(-\xi+\sqrt{-\Delta(k)})/2>0$ (the second mode is damped exponentially rapidly with a damping rate $\omega_i=(-\xi-\sqrt{-\Delta(k)})/2<0$).

The dispersion relation (\ref{dr1}) has been studied in detail for specific potentials of interaction (self-gravitating systems, chemotaxis, plasmas, and BMF model) in \cite{jeanschemo,paper5}.

\subsection{The response to a pulse}
\label{sec_pe}

The response of the system to a pulse is given by Eq. (\ref{pr15}). Using Eq. (\ref{lde9}), the evolution of the perturbation can be written as
\begin{equation}
\label{pe3}
\delta\hat{\Phi}({k},t)=(2\pi)^d\hat{u}(k) k^2 \rho I(k,t),
\end{equation}
with  
\begin{equation}
\label{pe4}
I(k,t)=\int_{\cal C}\frac{d\omega}{2\pi}e^{-i\omega t}\frac{1}{\omega^2+i\xi \omega-\omega_0^2(k)}.
\end{equation}
We recall that these equations determine the response function $R({k},t)$ [see Eq. (\ref{pr14})]. The integral (\ref{pe4}) can be easily calculated with the residue theorem.

\subsubsection{Stable case}

We first consider the stable case $\omega_0^2(k)\ge 0$.  For the Euler equation,
\begin{equation}
\label{pe5}
I(k,t)=-\frac{\sin\lbrack \omega_0(k)t\rbrack}{\omega_0(k)}.
\end{equation}
For the Smoluchowski equation,
\begin{equation}
\label{pe6}
I(k,t)=-\frac{1}{\xi}e^{-\frac{\omega_0^2(k)}{\xi}t}.
\end{equation}
For the damped Euler equation,
\begin{equation}
\label{pe7}
I(k,t)=-\frac{2}{\sqrt{\Delta(k)}}e^{-\xi t/2}\sin\left (\frac{1}{2}\sqrt{\Delta(k)}t\right ), \qquad (\Delta(k)>0),
\end{equation}
\begin{equation}
\label{pe8}
I(k,t)=-\frac{2}{\sqrt{-\Delta(k)}}e^{-\xi t/2}\sinh\left (\frac{1}{2}\sqrt{-\Delta(k)}t\right ), \qquad (\Delta(k)<0),
\end{equation}
\begin{equation}
\label{pe9}
I(k,t)=-t e^{-\xi t/2}, \qquad (\Delta(k)=0).
\end{equation}
The evolution of the perturbation is consistent with the discussion given in Section \ref{sec_dr}.

\subsubsection{Unstable case}
\label{sec_unstable1}

We now consider the unstable case $\omega_0^2(k)\le 0$ and define $\gamma_0^2(k)=-\omega_0^2(k)$. We can either compute the integral (\ref{pe4}) with the residue theorem or replace $\omega_0(k)$ by $i\gamma_0(k)$ in Eqs. (\ref{pe5})-(\ref{pe9}). For the Euler equation,
\begin{equation}
\label{pe10}
I(k,t)=-\frac{\sinh\lbrack \gamma_0(k)t\rbrack}{\gamma_0(k)}.
\end{equation}
For the Smoluchowski equation,
\begin{equation}
\label{pe11}
I(k,t)=-\frac{1}{\xi}e^{\frac{\gamma_0^2(k)}{\xi}t}.
\end{equation}
For the damped Euler equation,
\begin{equation}
\label{pe12}
I(k,t)=-\frac{2}{\sqrt{-\Delta(k)}}e^{-\xi t/2}\sinh\left (\frac{1}{2}\sqrt{-\Delta(k)}t\right ),
\end{equation}
with $\Delta(k)=-\xi^2-4\gamma_0^2(k)<0$.
The evolution of the perturbation is consistent with the discussion given in Section \ref{sec_dr}.

\subsection{The response to a step function}

The response of the system to a step function is given by Eq. (\ref{pr20}).  Using Eq. (\ref{lde9}), the evolution of the perturbation can be written as
\begin{equation}
\label{se2}
\delta\hat{\Phi}({\bf k},t)=(2\pi)^d\hat{u}(k) k^2 \rho J(k,t) \hat{\Phi}_e({\bf k}),
\end{equation}
with
\begin{equation}
\label{se3}
J(k,t)=\int_{\cal C}\frac{d\omega}{2\pi}e^{-i\omega t}\frac{1}{\omega^2+i\xi \omega-\omega_0^2(k)}\frac{i}{\omega}.
\end{equation}
The integral (\ref{se3}) can be easily calculated with the residue theorem.

\subsubsection{Stable case}
\label{sec_mar}

We first consider the stable case $\omega_0^2(k)\ge 0$. For the Euler equation,
\begin{equation}
\label{se4}
J(k,t)=-\frac{1}{\omega_0^2(k)}(1-\cos\lbrack \omega_0(k)t\rbrack).
\end{equation}
For the Smoluchowski equation,
\begin{equation}
\label{se5}
J(k,t)=-\frac{1}{\omega_0^2(k)}\left (1-e^{-\frac{\omega_0^2(k)}{\xi}t}\right ).
\end{equation}
For the damped Euler equation,
\begin{equation}
\label{se6}
J(k,t)=-\frac{1}{\omega_0^2(k)}+\frac{1}{\omega_0^2(k)}e^{-\xi t/2}\left\lbrack \frac{\xi}{\sqrt{\Delta(k)}}\sin \left (\frac{\sqrt{\Delta(k)}}{2}t\right )+\cos\left (\frac{\sqrt{\Delta(k)}}{2}t\right )\right\rbrack,\qquad (\Delta(k)>0),
\end{equation}
\begin{equation}
\label{se7}
J(k,t)=-\frac{1}{\omega_0^2(k)}+\frac{1}{\omega_0^2(k)}e^{-\xi t/2}\left\lbrack \frac{\xi}{\sqrt{-\Delta(k)}}\sinh \left (\frac{\sqrt{-\Delta(k)}}{2}t\right )+\cosh\left (\frac{\sqrt{-\Delta(k)}}{2}t\right )\right\rbrack,\qquad (\Delta(k)<0),
\end{equation}
\begin{equation}
\label{se8}
J(k,t)=-\frac{4}{\xi^2}+\frac{4}{\xi^2}e^{-\xi t/2}\left (\frac{1}{2}\xi t+1\right ),\qquad (\Delta(k)=0).
\end{equation}

The pole at $\omega=0$ contributes to the integral $J(k,t)$ by a constant term $1/\omega_0^2(k)$. When $\xi=0$, the zeros of the dielectric function $\epsilon(k,\omega)$ lie on the real axis ($\omega_r=\pm\omega_0(k)$, $\omega_i=0$). As a result, the perturbation oscillates indefinitely about $1/\omega_0^2(k)$ with a pulsation $\pm\omega_0(k)$ [see Eq. (\ref{se4})]. When $\xi>0$, the zeros of the dielectric function $\epsilon(k,\omega)$ strictly lie on the lower half plane ($\omega_i<0$). In that case, the perturbation asymptotically tends to $1/\omega_0^2(k)$ for $t\rightarrow +\infty$ [see Eqs. (\ref{se5})-(\ref{se8})].

\subsubsection{Unstable case}
\label{sec_unstable2}

We now consider the unstable case $\omega_0^2(k)\le 0$ and define $\gamma_0^2(k)=-\omega_0^2(k)$. We can either compute the integral (\ref{se3}) with the residue theorem or replace $\omega_0(k)$ by $i\gamma_0(k)$ in Eqs. (\ref{se4})-(\ref{se8}). For the Euler equation,
\begin{equation}
\label{se9}
J(k,t)=\frac{1}{\gamma_0^2(k)}(1-\cosh\lbrack \gamma_0(k)t\rbrack).
\end{equation}
For the Smoluchowski equation,
\begin{equation}
\label{se10}
J(k,t)=\frac{1}{\gamma_0^2(k)}\left (1-e^{\frac{\gamma_0^2(k)}{\xi}t}\right ).
\end{equation}
For the damped Euler equation,
\begin{equation}
\label{se11}
J(k,t)=\frac{1}{\gamma_0^2(k)}-\frac{1}{\gamma_0^2(k)}e^{-\xi t/2}\left\lbrack \frac{\xi}{\sqrt{-\Delta(k)}}\sinh \left (\frac{\sqrt{-\Delta(k)}}{2}t\right )+\cosh\left (\frac{\sqrt{-\Delta(k)}}{2}t\right )\right\rbrack,
\end{equation}
with $\Delta(k)=-\xi^2-4\gamma_0^2(k)<0$.

{\it Remark:} In the unstable case, the perturbation grows exponentially rapidly as explained in Section \ref{sec_dr}. Of course, the linear response theory ceases to be valid when the perturbation has grown significatively, so the expressions obtained in Sections \ref{sec_unstable1} and \ref{sec_unstable2} are only valid for sufficiently short times.

\subsection{The polarization function}

The polarization function is given by Eq. (\ref{lde8}). It can be  written as
\begin{equation}
\label{pfe2}
P({k},t)=(2\pi)^d\hat{u}(k) k^2 \rho K(k,t),
\end{equation}
with
\begin{equation}
\label{pfe3}
K(k,t)=\int_{\cal C}\frac{d\omega}{2\pi}e^{-i\omega t}\frac{1}{\omega^2+i\xi \omega-c_s^2 k^2}.
\end{equation}
Comparing the polarization function (\ref{lde8}) with the response
function (\ref{lde7}), we see that they only differ by the replacement
of $\omega_0^2(k)$ by $c_s^2k^2$ (this amounts to neglecting the
self-interaction as indicated in Section \ref{sec_pr}). We can therefore readily adapt the results of
Section \ref{sec_pe}. For the Euler equation,
\begin{equation}
\label{pfe4}
K(k,t)=-\frac{\sin (c_s k t)}{c_s k}.
\end{equation}
For the Smoluchowski equation,
\begin{equation}
\label{pfe5}
K(k,t)=-\frac{1}{\xi}e^{-\frac{c_s^2 k^2}{\xi}t}.
\end{equation}
For the damped Euler equation,
\begin{equation}
\label{pfe6}
K(k,t)=-\frac{2}{\sqrt{\Delta(k)}}e^{-\xi t/2}\sin\left (\frac{1}{2}\sqrt{\Delta(k)}t\right ), \qquad (\Delta(k)>0),
\end{equation}
\begin{equation}
\label{pfe7}
K(k,t)=-\frac{2}{\sqrt{-\Delta(k)}}e^{-\xi t/2}\sinh\left (\frac{1}{2}\sqrt{-\Delta(k)}t\right ), \qquad (\Delta(k)<0),
\end{equation}
\begin{equation}
\label{pfe8}
K(k,t)=-t e^{-\xi t/2}, \qquad (\Delta(k)=0),
\end{equation}
with $\Delta(k)=4 c_s^2 k^2-\xi^2$.

\section{The Vlasov equation}
\label{sec_vlasov}

We consider a system of particles with long-range interactions interacting via a binary potential $u(|{\bf r}-{\bf r}'|)$. We assume that ``collisions'' (correlations, graininess, finite $N$ effects) are negligible. In that case, the system is described by the Vlasov equation
\begin{equation}
\label{v1}
\frac{\partial{f}}{\partial t}+{\bf v}\cdot \frac{\partial {f}}{\partial {\bf r}}-(\nabla\Phi+\nabla\Phi_{e})\cdot \frac{\partial f}{\partial {\bf v}}=0,
\end{equation}
\begin{equation}
\label{v2}
\Phi({\bf r},t)=\int u(|{\bf r}-{\bf r}'|) \rho({\bf r}',t)\, d{\bf r}',
\end{equation}
where $f({\bf r},{\bf v},t)$ is the distribution function and $\rho({\bf r},t)=\int f({\bf r},{\bf v},t)\, d{\bf v}$ the density. As before, we have distinguished the potential $\Phi({\bf r},t)$ produced by the particles from the external potential $\Phi_e({\bf r},t)$. The Vlasov equation, which is based on a mean field approximation, describes the collisionless evolution of stellar systems, plasmas, and of the HMF model. It is rigorously valid for systems with long-range interaction in a proper  thermodynamic limit $N\rightarrow +\infty$ \cite{bh}. Systems with long-range interactions are known to organize spontaneously into quasi stationary states (QSSs) that are steady states of the Vlasov equation \cite{bt,cdr}. Galaxies in astrophysics and large-scale vortices in 2D hydrodynamics are examples of such QSSs \cite{houchesPH}. These QSSs have also been studied extensively for toy models like the HMF model \cite{ar,latora,lrt,yamaguchi,epjblb,precommun,prl1,prl2,campa1,campa2,bachelard1,
bachelard2,baldovin,staniscia1,barre1,staniscia2,barre2,oy,firpo,levin,rocha} or simplified models of gravitational dynamics \cite{hohl68,goldstein69,cuperman69,lecar71,janin71,tanekusa87,mineau,
yamaguchi2008,levingrav,levin2D,gabrielli2010,jw2011,tlp}. The linear response theory may be applied to systems in such QSSs. For simplicity, we restrict ourselves to spatially homogeneous systems as in \cite{patelli}. The case of spatially inhomogeneous systems can be treated with angle-action variables as in \cite{oyresponse}.

\subsection{The linearized Vlasov equation}
\label{sec_vl}

We consider a system in a steady state with uniform density $\rho$ and distribution function $f({\bf v})$, and examine its response to a small external potential $\Phi_e({\bf r},t)$. Since the perturbation is small, we can develop a linear response theory. The linearized Vlasov equation is
\begin{equation}
\label{vl1}
\frac{\partial\delta{f}}{\partial t}+{\bf v}\cdot \frac{\partial \delta {f}}{\partial {\bf r}}-(\nabla\delta\Phi+\nabla\Phi_{e})\cdot \frac{\partial f}{\partial {\bf v}}=0,
\end{equation}
\begin{equation}
\label{vl2}
\delta\Phi({\bf r},t)=\int u(|{\bf r}-{\bf r}'|)\delta \rho({\bf r}',t)\, d{\bf r}'.
\end{equation}
Taking the Fourier-Laplace transform of this equation, and assuming $\delta\tilde f(t=0)=0$, we get
\begin{equation}
\label{vl3}
\delta\tilde{f}({\bf k},{\bf v},\omega)=\frac{{\bf k}\cdot \frac{\partial f}{\partial {\bf v}}}{{\bf k}\cdot {\bf v}-\omega}\left\lbrack \delta\tilde\Phi({\bf k},\omega)+\tilde\Phi_e({\bf k},\omega)\right\rbrack,
\end{equation}
\begin{equation}
\label{vl4}
\delta\tilde\Phi({\bf k},\omega)=(2\pi)^d\hat{u}(k)\delta \tilde\rho({\bf k},\omega).
\end{equation}
Integrating Eq. (\ref{vl3}) over the velocity, and using Eq. (\ref{vl4}), we find that the Fourier-Laplace transform of the perturbed potential and of the perturbed distribution function are given by
\begin{equation}
\label{vl3bma}
\delta\tilde{\Phi}({\bf k},\omega)=\frac{1-\epsilon({\bf k},\omega)}{\epsilon({\bf k},\omega)}\tilde\Phi_e({\bf k},\omega),
\end{equation}
\begin{equation}
\label{vl3b}
\delta\tilde{f}({\bf k},{\bf v},\omega)=\frac{{\bf k}\cdot \frac{\partial f}{\partial {\bf v}}}{{\bf k}\cdot {\bf v}-\omega}\frac{1}{\epsilon({\bf k},\omega)}\tilde\Phi_e({\bf k},\omega),
\end{equation}
with the dielectric function
\begin{equation}
\label{vl5}
\epsilon({\bf k},\omega)=1-(2\pi)^d\hat{u}(k)\int \frac{{\bf k}\cdot \frac{\partial f}{\partial {\bf v}}}{{\bf k}\cdot {\bf v}-\omega}\, d{\bf v}.
\end{equation}
We consider a $d$-dimensional space. It is convenient to take the $v_d$-axis in the direction of ${\bf k}$. If we integrate over $v_1...v_d$ and note $v$ for $v_d$ and $f(v)$ for $\int f\, dv_1...dv_{d-1}$ (in the following, $f(v)$ will be called the reduced distribution function), we obtain
\begin{equation}
\label{vl6}
\epsilon(k,\omega)=1-(2\pi)^d\hat{u}(k)\int_{L} \frac{f'(v)}{v-\omega/k}\, d{v},
\end{equation}
where the integration has to be performed along the Landau contour $L$ \cite{landau}. For the Maxwell distribution (\ref{app1}), we can write the dielectric function in the form
\begin{equation}
\label{vl7}
\epsilon(k,\omega)=1+(2\pi)^d\hat{u}(k)\beta \rho W\left (\sqrt{\beta}\frac{\omega}{k}\right),
\end{equation}
where
\begin{equation}
\label{vl8}
W(z)=\frac{1}{\sqrt{2\pi}}\int_{L}\frac{x}{x-z}e^{-x^2/2}\, dx,
\end{equation}
is the plasma dispersion function \cite{fried}.

\subsection{The dispersion relation}
\label{sec_drv}

The dispersion relation $\epsilon({k},\omega)=0$ can be written explicitly as
\begin{equation}
\label{drv1}
1-(2\pi)^d\hat{u}(k)\int_{L} \frac{f'(v)}{v-\omega/k}\, d{v}=0.
\end{equation}
It is in general difficult to solve this equation analytically except for the waterbag and the Cauchy distributions (see Sections \ref{sec_waterbag} and \ref{sec_cauchy}). Analytical results can be obtained  for the Maxwellian distribution in some asymptotic limits \cite{balescubook,bt,nyquist}. In general, the imaginary part of the complex pulsation $\omega$ is non-zero, leading to Landau damping ($\omega_i<0$) or Landau growth  ($\omega_i>0$). Using the Nyquist theorem \cite{nyquist}, it is possible to obtain a general criterion of dynamical stability. If the distribution function $f(v)$ has a single maximum at $v=0$, it can be shown that the  system is stable with respect to a perturbation with wavenumber $k$ when
\begin{equation}
\label{drv2}
\epsilon(k,0)=1-(2\pi)^d\hat{u}(k)\int_{-\infty}^{+\infty} \frac{f'(v)}{v}\, d{v}>0,
\end{equation}
and unstable otherwise. This is a generalization of the Jeans stability criterion in astrophysics \cite{bt}. For repulsive potentials $\hat{u}(k)>0$ (as in plasma physics), a single humped distribution is always stable. For attractive potentials $\hat{u}(k)<0$ (as in astrophysics), a single humped distribution  may be unstable to some wavelengths. When the distribution function is of the form $f=F(v^2/2)$, using the notion of ``corresponding barotropic gas'' (see Appendix \ref{sec_corr}), one can show that the stability criterion (\ref{drv2}) can be written as
\begin{equation}
\label{drv3}
\epsilon(k,0)=1+\frac{(2\pi)^d\hat{u}(k)\rho}{c_s^2} > 0,
\end{equation}
where $c_s$ is the velocity of sound in the corresponding barotropic gas. The equivalence between Eqs. (\ref{drv2}) and (\ref{drv3}) is due to the identity (\ref{corr3}). On the other hand, Eq. (\ref{drv3}) is equivalent to the stability condition (\ref{dr5}) for a collision-dominated gas. Therefore, for spatially homogeneous distributions, a collisionless system with a distribution function $f=F(v^2/2)$ is stable if, and only, if the corresponding barotropic gas is stable \cite{nyquist}. This equivalence is not true anymore for spatially inhomogeneous distributions \cite{bt,aaantonov}.

\subsection{The polarization function}

The Fourier-Laplace transform of the polarization function is related to the dielectric function by Eq. (\ref{pr10}). Using Eq. (\ref{vl5}), we obtain
\begin{equation}
\label{pol1}
P({\bf k},\omega)=(2\pi)^d\hat{u}(k)\int\frac{{\bf k}\cdot \frac{\partial f}{\partial {\bf v}}}{{\bf k}\cdot {\bf v}-\omega}\, d{\bf v}.
\end{equation}
Taking the inverse Laplace transform of this equation, we get
\begin{equation}
\label{pol2}
P({\bf k},t)=(2\pi)^d\hat{u}(k)i\int {\bf k}\cdot \frac{\partial f}{\partial {\bf v}}e^{-i{\bf k}\cdot {\bf v}t}\, d{\bf v}.
\end{equation}
For the Maxwell distribution (\ref{app1}), we have the explicit result
\begin{equation}
\label{pol4}
P({k},t)=-(2\pi)^d\hat{u}(k)k^2 \rho\,  t \, e^{-\frac{k^2 t^2}{2\beta}}.
\end{equation}
The polarization function of a collisionless system with a Maxwell distribution decays to zero, contrary to the polarization of an ideal fluid which oscillates [see Eq. (\ref{pfe4})]. This is a consequence of phase mixing \cite{bt}.

\subsection{An integral equation for the response function}

The Fourier transform of the linearized Vlasov equation (\ref{vl1}) is
\begin{equation}
\label{ir1}
\frac{\partial\delta\hat{f}}{\partial t}+i{\bf k}\cdot {\bf v}\delta\hat{f}-i{\bf k}\cdot \frac{\partial f}{\partial {\bf v}}\left\lbrack \delta\hat{\Phi}({\bf k},t)+\hat{\Phi}_{e}({\bf k},t)\right \rbrack=0.
\end{equation}
Assuming $\delta {\hat f}(t=0)=0$, Eq. (\ref{ir1}) can be integrated into
\begin{equation}
\label{ir2}
\delta\hat{f}({\bf k},{\bf v},t)=i{\bf k}\cdot \frac{\partial f}{\partial {\bf v}}\int_0^t \, dt'\left\lbrack \delta\hat{\Phi}({\bf k},t')+\hat{\Phi}_{e}({\bf k},t')\right \rbrack e^{-i{\bf k}\cdot {\bf v}(t-t')},
\end{equation}
where we recall that $\delta\hat{\Phi}({\bf k},t)$ depends implicitly on $\delta\hat{f}({\bf k},{\bf v},t)$ through Eq. (\ref{vl4}). Integrating Eq. (\ref{ir2}) over the velocity, we get
\begin{equation}
\label{ir3}
\delta\hat{\rho}({\bf k},t)=i\int d{\bf v}\, {\bf k}\cdot \frac{\partial f}{\partial {\bf v}}\int_0^t \, dt'\left\lbrack \delta\hat{\Phi}({\bf k},t')+\hat{\Phi}_{e}({\bf k},t')\right \rbrack e^{-i{\bf k}\cdot {\bf v}(t-t')}.
\end{equation}
Using Eq. (\ref{vl4}), we find that the perturbed potential satisfies an equation of the form
\begin{equation}
\label{ir4}
\delta\hat{\Phi}({\bf k},t)=(2\pi)^d\hat{u}(k)i\int d{\bf v}\, {\bf k}\cdot \frac{\partial f}{\partial {\bf v}}\int_0^t \, dt'\left\lbrack \delta\hat{\Phi}({\bf k},t')+\hat{\Phi}_{e}({\bf k},t')\right \rbrack e^{-i{\bf k}\cdot {\bf v}(t-t')}.
\end{equation}
Comparing Eq. (\ref{ir4}) with Eq. (\ref{pr6}), we recover the expression (\ref{pol2}) of the polarization function. On the other hand, for a pulse $\hat{\Phi}_{e}({\bf k},t')=\delta(t')$, the perturbed potential $\delta\hat{\Phi}({\bf k},t)$ is equal to the response function $R({\bf k},t)$ [see Eq. (\ref{pr14})]. Therefore, the response function satisfies the integral equation
\begin{equation}
\label{ir5}
R({\bf k},t)=(2\pi)^d\hat{u}(k)i\int d{\bf v}\, {\bf k}\cdot \frac{\partial f}{\partial {\bf v}} e^{-i{\bf k}\cdot {\bf v}t}+(2\pi)^d\hat{u}(k)i\int d{\bf v}\, {\bf k}\cdot \frac{\partial f}{\partial {\bf v}}\int_0^t \, dt' e^{-i{\bf k}\cdot {\bf v}(t-t')} R({\bf k},t').
\end{equation}
For the Maxwellian distribution, we can reduce the foregoing equation to the form
\begin{equation}
\label{ir6}
R({k},t)=-(2\pi)^d\hat{u}(k)k^2\rho t  e^{-\frac{k^2 t^2}{2\beta}}-(2\pi)^d\hat{u}(k)k^2\rho \int_0^t \, dt' (t-t') e^{-\frac{k^2}{2\beta}(t-t')^2} R({k},t').
\end{equation}
The first term accounts for phase mixing [see Eq. (\ref{pol4})] and the second term for Landau damping or Landau growth \cite{bt}.

\subsection{The waterbag distribution}
\label{sec_waterbag}

We consider a reduced distribution function of the form
$f(v)=\rho/2v_m$ for $-v_m\le v\le v_m$ and $f(v)=0$ otherwise (we note that  $f(v)\rightarrow \rho\delta(v)$ when $v_m\rightarrow 0$). This is the so-called waterbag distribution. Using $f'(v)=(\rho/2v_m)\lbrack \delta(v+v_m)-\delta(v-v_m)\rbrack$, the dielectric function (\ref{vl6}) is explicitly given by
\begin{equation}
\label{w1}
\epsilon(k,\omega)=1-\frac{(2\pi)^d\hat{u}(k)k^2\rho}{\omega^2-v_m^2k^2}.
\end{equation}
The dispersion relation $\epsilon(k,\omega)=0$ can be written as
\begin{equation}
\label{w2}
\omega^2=v_m^2 k^2+(2\pi)^d\hat{u}(k)k^2\rho.
\end{equation}
The pulsation is purely real or purely imaginary. The system is stable with respect to a perturbation with wavenumber $k$ when
\begin{equation}
\label{w3}
\epsilon(k,0)=1+\frac{(2\pi)^d\hat{u}(k)\rho}{v_m^2}>0,
\end{equation}
and unstable otherwise.  In the stable case, the perturbation oscillates with a pulsation $\omega_r=\pm \sqrt{\omega^2}$ without being damped. In the unstable case, the perturbation grows exponentially rapidly with a rate $\omega_i=\sqrt{-\omega^2}>0$ (the second mode is damped exponentially rapidly with a rate $\omega_i=-\sqrt{-\omega^2}<0$).

The important point to notice is that the dielectric function (\ref{w1}) of a collisionless system described by the Vlasov equation with the waterbag distribution  coincides with the dielectric function (\ref{lde9}) of a collision-dominated gas described by the Euler equation ($\xi=0$), provided that the velocity of sound $c_s$ is replaced by the maximum velocity $v_m$. Therefore, the results obtained in Section \ref{sec_de} for the Euler equation are immediately applicable to the Vlasov equation when the unperturbed system is described by the waterbag distribution. A property of the waterbag distribution is that it does not experience Landau damping so the perturbation has a purely oscillatory behaviour in the stable case as for a perfect gas [see Eqs. (\ref{pe5}), (\ref{se4}), and (\ref{pfe4})].

\section{The Cauchy distribution}
\label{sec_cauchy}

It is in general difficult to solve the dispersion relation (\ref{drv1}) of the linearized Vlasov equation analytically, even for the Maxwell distribution. Analytical results can be obtained for the waterbag distribution (see Section \ref{sec_waterbag}), but this distribution is very particular because the zeros of the dielectric function lie on the real axis (in the stable case) so the perturbations do not experience Landau damping. This is to be contrasted with most distribution functions, including the Maxwellian. In this respect, it can be interesting to consider the Cauchy distribution
\begin{equation}
\label{c1}
f(v)=\frac{\rho}{\pi u_0}\frac{1}{1+\frac{v^2}{u_0^2}},
\end{equation}
which is less peculiar than the waterbag distribution, and for which analytical results can be obtained (a drawback of this distribution is that the mean square velocity diverges). 

\subsection{Dispersion relation}
\label{sec_cauchydr}

For the Cauchy distribution, the dielectric function (\ref{vl6}) can be written as
\begin{equation}
\label{c2}
\epsilon(k,\omega)=1+(2\pi)^d\hat{u}(k)\frac{2\rho}{\pi u_0^2}\int_L \frac{x}{(1+x^2)^2(x-\frac{\omega}{k u_0})}\, dx.
\end{equation}
For $\omega$ lying in the upper half plane, the integral can be computed analytically by adding to the real axis a large semi-circle in the lower half-plane and using the Cauchy residue theorem for a function with a double pole at $x=-i$. The evaluation of the integral leads to the result 
\begin{equation}
\label{c3}
\epsilon(k,\omega)=1-\frac{(2\pi)^d\hat{u}(k)\rho k^2}{(i k u_0+{\omega})^2}.
\end{equation}
This result is then extended to any $\omega$ by analytic continuation. The dispersion relation $\epsilon(k,\omega)=0$ is readily solved, giving the two complex roots:
\begin{equation}
\label{c4}
\omega=\pm\sqrt{(2\pi)^d\hat{u}(k)k^2\rho}-i u_0 k.
\end{equation}

For a repulsive potential $\hat{u}(k)>0$, the system is stable. The perturbation oscillates with a pulsation $\omega_r=\pm \sqrt{(2\pi)^d\hat{u}(k)k^2\rho}$ and is damped exponentially rapidly with a rate $\omega_i=-u_0 k<0$. For an attractive potential $\hat{u}(k)<0$, the system is stable with respect to a perturbation with wavenumber $k$ when
\begin{equation}
\label{c7}
\epsilon(k,0)=1+\frac{(2\pi)^d\hat{u}(k)\rho}{u_0^2}>0,
\end{equation}
and unstable otherwise. In the stable case, the perturbation is damped exponentially rapidly with a rate $\omega_i=\sqrt{(2\pi)^d|\hat{u}(k)|k^2\rho}-u_0 k<0$. In the unstable case, the perturbation grows exponentially rapidly with a rate $\omega_i=\sqrt{(2\pi)^d|\hat{u}(k)|k^2\rho}-u_0 k>0$. The other mode is always damped exponentially rapidly with a rate $\omega_i=-\sqrt{(2\pi)^d|\hat{u}(k)|k^2\rho}-u_0 k<0$.

We note that  the dielectric function (\ref{c3}) of a collisionless system described by the Vlasov equation with the Cauchy distribution is similar to the dielectric function (\ref{lde9}) of a collision-dominated gas described by the damped Euler equations ($\xi>0$), where $u_0$ plays the role of $c_s$ and $2ku_0$ the role of $\xi$. They are, however, different because the friction coefficient $\xi$ in Eq. (\ref{lde9}) is constant while the damping term $2 k u_0$ in Eq. (\ref{c3}) depends on the wavenumber $k$  (they coincide only for the HMF model for which $k=1$).

\subsection{Polarization function}

The Fourier-Laplace transform of the polarization function is related to the dielectric function by Eq. (\ref{pr10}). Using Eq. (\ref{c3}), we get
\begin{equation}
\label{c8}
P(k,\omega)=\frac{(2\pi)^d\hat{u}(k)\rho k^2}{(i k u_0+{\omega})^2}.
\end{equation}
Taking the inverse Laplace transform of Eq. (\ref{c8}), we obtain the explicit result
\begin{equation}
\label{c9}
P(k,t)=-(2\pi)^d\hat{u}(k)\rho k^2 t e^{-u_0 kt}.
\end{equation}
The polarization function decreases exponentially rapidly as a result of phase mixing (we note that the polarization function (\ref{pol4}) associated with the Maxwellian distribution has a Gaussian decay).

\subsection{The response to a pulse}

The Fourier-Laplace transform of the response function is related to the dielectric function by Eq. (\ref{pr11}). Using Eq. (\ref{c3}), we get
\begin{equation}
\label{c10}
R(k,\omega)=\frac{(2\pi)^d\hat{u}(k)k^2{\rho}}{(i k u_0+{\omega})^2-(2\pi)^d\hat{u}(k)k^2{\rho}}.
\end{equation}
The response of the system to a pulse is given by Eq. (\ref{pr14}) with Eq. (\ref{c10}). Taking its inverse Laplace transform, we obtain
\begin{equation}
\label{c11}
\delta\hat{\Phi}({k},t)=\int_{\cal C}\frac{d\omega}{2\pi}e^{-i\omega t}\frac{(2\pi)^d\hat{u}(k)k^2{\rho}}{(\omega+i u_0 k+\sqrt{(2\pi)^d\hat{u}(k)k^2\rho})(\omega+i u_0 k-\sqrt{(2\pi)^d\hat{u}(k)k^2\rho})}.
\end{equation}
This integral can be easily performed with the residue theorem. For a  repulsive potential ($\hat{u}(k)>0$),
\begin{equation}
\label{c12}
\delta\hat{\Phi}({k},t)=-\sqrt{(2\pi)^d\hat{u}(k)k^2{\rho}}\sin\left (\sqrt{(2\pi)^d\hat{u}(k)k^2\rho}\, t\right ) e^{-u_0 kt},
\end{equation}
and for an attractive potential ($\hat{u}(k)<0$),
\begin{equation}
\label{c13}
\delta\hat{\Phi}({k},t)=\sqrt{(2\pi)^d|\hat{u}(k)|k^2{\rho}}\sinh\left (\sqrt{(2\pi)^d|\hat{u}(k)|k^2\rho}\, t\right ) e^{-u_0 kt}.
\end{equation}
The evolution of the perturbation is consistent with the discussion given in Section \ref{sec_cauchydr}.

\subsection{The response to a step function}

The response of the system to a step function is given by Eq. (\ref{pr19}) with Eq. (\ref{c10}). Taking its inverse Laplace transform,  we obtain
\begin{equation}
\label{c14}
\delta\hat{\Phi}({\bf k},t)=i\hat{\Phi}_{e}({\bf k})\int_{\cal C}\frac{d\omega}{2\pi}e^{-i\omega t}\frac{(2\pi)^d\hat{u}(k)k^2{\rho}}{\omega (\omega+i u_0 k+\sqrt{(2\pi)^d\hat{u}(k)k^2\rho})(\omega+i u_0 k-\sqrt{(2\pi)^d\hat{u}(k)k^2\rho})}.
\end{equation}
This integral can be easily performed with the residue theorem. For a  repulsive potential ($\hat{u}(k)>0$),
\begin{equation}
\label{c15}
\delta\hat{\Phi}({\bf k},t)=\delta\hat{\Phi}_{\infty}({\bf k})\left\lbrace 1-e^{-u_0 kt}\left\lbrack \sqrt{\frac{u_0^2}{(2\pi)^d\hat{u}(k)\rho}}\sin\left (\sqrt{(2\pi)^d\hat{u}(k)k^2\rho}\, t\right )+\cos\left (\sqrt{(2\pi)^d\hat{u}(k)k^2\rho}\, t\right )\right\rbrack\right\rbrace,
\end{equation}
and for an attractive potential ($\hat{u}(k)<0$),
\begin{equation}
\label{c16}
\delta\hat{\Phi}({\bf k},t)=\delta\hat{\Phi}_{\infty}({\bf k})\left\lbrace 1-e^{-u_0 kt}\left\lbrack \sqrt{\frac{u_0^2}{(2\pi)^d|\hat{u}(k)|\rho}}\sinh\left (\sqrt{(2\pi)^d|\hat{u}(k)|k^2\rho}\, t\right )+\cosh\left (\sqrt{(2\pi)^d|\hat{u}(k)|k^2\rho}\, t\right )\right\rbrack\right\rbrace,
\end{equation}
where
\begin{equation}
\label{c17}
\delta\hat{\Phi}_{\infty}({\bf k})=-\frac{(2\pi)^d\hat{u}(k)\rho k^2}{u_0^2 k^2+(2\pi)^d\hat{u}(k)k^2\rho}\hat{\Phi}_{e}({\bf k}).
\end{equation}
In the stable case, the zeros of the dielectric function lie in the lower half-plane and the perturbation asymptotically tends towards the distribution defined by Eq. (\ref{c17}).

\section{Linearized hydrodynamic equations}
\label{sec_hydro}

In this section, we derive hydrodynamic equations satisfied by the perturbed density $\delta\hat{\rho}({\bf k},t)$ in the linear regime. We consider the case of a collision-dominated gas (Euler) and the case of a collisionless system (Vlasov) and show the correspondence between these two systems.

\subsection{Linearized hydrodynamic equations for a collision-dominated gas}

For a collision-dominated gas, it is straightforward to obtain the equation for the perturbed density $\delta{\rho}({\bf r},t)$. Taking the time derivative of Eq. (\ref{lde1}) and the divergence of Eq. (\ref{lde2}), and combining the resulting equations, we obtain
\begin{equation}
\label{h1}
\frac{\partial^2\delta\rho}{\partial t^2}+\xi\frac{\partial\delta\rho}{\partial t}-c_s^2\Delta\delta\rho=\rho\Delta\delta\Phi+\rho\Delta\Phi_e.
\end{equation}
This equation without the right hand side is called the telegraph equation. However, Eq. (\ref{h1}) is more complicated than the telegraph equation because $\delta\Phi$ depends on $\delta\rho$ according to Eq. (\ref{lde3}). It is therefore an integro-differential equation. Taking the Fourier transform of Eq. (\ref{h1}), we get
\begin{equation}
\label{h2}
\frac{\partial^2\delta\hat\rho}{\partial t^2}+\xi\frac{\partial\delta\hat\rho}{\partial t}+c_s^2 k^2\delta\hat\rho=-\rho k^2 (\delta\hat\Phi+\hat\Phi_e),
\end{equation}
where $\delta\hat\Phi$ is related to $\delta\hat\rho$ by Eq. (\ref{lde6}). The foregoing equation can be rewritten as
\begin{equation}
\label{h3}
\frac{\partial^2\delta\hat\rho}{\partial t^2}+\xi\frac{\partial\delta\hat\rho}{\partial t}+\omega_0^2(k)\delta\hat\rho=-\rho k^2 \hat\Phi_e.
\end{equation}
For $\xi=0$ (Euler), it reduces to
\begin{equation}
\label{h4}
\frac{\partial^2\delta\hat\rho}{\partial t^2}+\omega_0^2(k)\delta\hat\rho=-\rho k^2 \hat\Phi_e.
\end{equation}
For $\xi\rightarrow +\infty$ (Smoluchowski), we get
\begin{equation}
\label{h5}
\xi\frac{\partial\delta\hat\rho}{\partial t}+\omega_0^2(k)\delta\hat\rho=-\rho k^2 \hat\Phi_e.
\end{equation}

\subsection{Linearized hydrodynamic equations for a collisionless system}

For a collisionless system, the perturbed density $\delta\hat{\rho}({\bf k},t)$ satisfies the integral equation [see Eq. (\ref{ir3})]:
\begin{equation}
\label{h6}
\delta\hat{\rho}({\bf k},t)=i\int_0^t \, dt'\left\lbrack \delta\hat{\Phi}({\bf k},t')+\hat{\Phi}_{e}({\bf k},t')\right \rbrack\int d{\bf v}\, {\bf k}\cdot \frac{\partial f}{\partial {\bf v}} e^{-i{\bf k}\cdot {\bf v}(t-t')},
\end{equation}
where $\delta\hat{\Phi}({\bf k},t')$ is related to $\delta\hat{\rho}({\bf k},t')$ by Eq. (\ref{vl4}). Integrating the last term by parts, we obtain
\begin{equation}
\label{h7}
\delta\hat{\rho}({\bf k},t)=-k^2\int_0^t \, dt'\left\lbrack \delta\hat{\Phi}({\bf k},t')+\hat{\Phi}_{e}({\bf k},t')\right \rbrack (t-t')\int d{\bf v}\, f({\bf v}) e^{-i{\bf k}\cdot {\bf v}(t-t')}.
\end{equation}
Introducing the reduced distribution function $f(v)$, we get
\begin{equation}
\label{h8}
\delta\hat{\rho}({\bf k},t)=-k^2\int_0^t \, dt'\left\lbrack \delta\hat{\Phi}({\bf k},t')+\hat{\Phi}_{e}({\bf k},t')\right \rbrack (t-t')\int_{-\infty}^{+\infty} d{v}\, f({v}) e^{-i{k} {v}(t-t')}.
\end{equation}
Defining the Fourier transform  of $f(v)$ in velocity space by
\begin{equation}
\label{h9}
\hat{f}(q)=\int_{-\infty}^{+\infty} \frac{d{v}}{2\pi}\, e^{-i q v} f({v}),
\end{equation}
we can write Eq. (\ref{h8}) in the form
\begin{equation}
\label{h11}
\delta\hat{\rho}({\bf k},t)=-2\pi k\int_0^t \, dt'\left\lbrack \delta\hat{\Phi}({\bf k},t')+\hat{\Phi}_{e}({\bf k},t')\right \rbrack q \hat{f}(q),
\end{equation}
where $q=k(t-t')$.  Taking the first and second derivatives of this equation with respect to time, we get
\begin{equation}
\label{h12}
\frac{\partial\delta\hat{\rho}}{\partial t}=-2\pi k^2\int_0^t \, dt'\left\lbrack \delta\hat{\Phi}({\bf k},t')+\hat{\Phi}_{e}({\bf k},t')\right \rbrack \frac{d}{dq}\lbrack q \hat{f}(q)\rbrack,
\end{equation}
\begin{equation}
\label{h13}
\frac{\partial^2\delta\hat{\rho}}{\partial t^2}=-\rho k^2 \left\lbrack \delta\hat{\Phi}({\bf k},t)+\hat{\Phi}_{e}({\bf k},t)\right \rbrack-2\pi k^3\int_0^t \, dt'\left\lbrack \delta\hat{\Phi}({\bf k},t')+\hat{\Phi}_{e}({\bf k},t')\right \rbrack \frac{d^2}{dq^2}\lbrack q \hat{f}(q)\rbrack.
\end{equation}
To obtain Eq. (\ref{h13}), we have used the fact that $(d/dq)[q\hat{f}(q)]=q\hat{f}'(q)+\hat{f}(q)$ is equal to $\hat{f}(0)=\rho/2\pi$ when $q=0$. In general, we cannot obtain a  closed partial differential equation for $\delta\hat{\rho}({\bf k},t)$, unlike in a collision-dominated gas. An exception concerns the waterbag and Cauchy distributions.

\subsubsection{The waterbag distribution}

Taking the Fourier transform of the waterbag distribution defined in Section \ref{sec_waterbag}, we obtain
\begin{equation}
\label{h14}
\hat{f}(q)=\frac{\rho}{2\pi}\frac{\sin(v_m q)}{v_m q}.
\end{equation}
Substituting the identity
\begin{equation}
\label{h15}
\frac{d^2}{dq^2}\lbrack q \hat{f}(q)\rbrack=-v_m^2 q\hat{f}(q).
\end{equation}
in Eq. (\ref{h13}), and using Eq. (\ref{h11}), we get
\begin{equation}
\label{h16}
\frac{\partial^2\delta\hat\rho}{\partial t^2}+v_m^2  k^2\delta\hat\rho=-\rho k^2 (\delta\hat\Phi+\hat\Phi_e).
\end{equation}
Comparing this equation with Eq. (\ref{h2}), we see that a collisionless system with the waterbag distribution behaves as a collision-dominated gas in which the velocity of sound $c_s$ is replaced by the maximum velocity $v_m$. This system does not experience Landau damping. We note that the dispersion relation (\ref{w2}) of the waterbag distribution can be obtained directly from the  hydrodynamic equation (\ref{h16}).

\subsubsection{The Cauchy distribution}

Taking the Fourier transform of the Cauchy distribution (\ref{c1}), we obtain
\begin{equation}
\label{h17}
\hat{f}(q)=\frac{\rho}{2\pi}e^{-u_0 q}.
\end{equation}
Substituting the identity
\begin{equation}
\label{h18}
\frac{d^2}{dq^2}\lbrack q \hat{f}(q)\rbrack=-2u_0\frac{d}{dq}\lbrack q \hat{f}(q)\rbrack-u_0^2 q\hat{f}(q),
\end{equation}
in Eq. (\ref{h13}), and using Eqs. (\ref{h11}) and (\ref{h12}), we obtain
\begin{equation}
\label{h19}
\frac{\partial^2\delta\hat\rho}{\partial t^2}+2u_0 k\frac{\partial\delta\hat\rho}{\partial t}+u_0^2  k^2\delta\hat\rho=-\rho k^2 (\delta\hat\Phi+\hat\Phi_e).
\end{equation}
Comparing this equation with Eq. (\ref{h2}), we see that a collisionless system with the Cauchy distribution behaves as a collision-dominated gas in which the velocity of sound $c_s$ is replaced by $u_0$. Furthermore, this equation exhibits a damping in Fourier space with a damping coefficient proportional to $k$. This is of course a manifestation of Landau damping. However, it appears here in a simple manner without having to carry out contour integration in the complex plane. We note that the dispersion relation (\ref{c4}) of the Cauchy distribution can be obtained directly from the  hydrodynamic equation (\ref{h19}).

\section{Asymptotic value of the perturbation}
\label{sec_asy}

The response of the system to a step function is given by the general expression (\ref{pr21}).  In the case where all the zeros of the dielectric function $\epsilon({\bf k},\omega)$ strictly lie in the lower half plane ($\omega_i<0$), an assumption that will be made in this section, the perturbation $\delta\hat\Phi({\bf k},t)$ relaxes towards the asymptotic value $\delta\hat{\Phi}_{\infty}({\bf k})$ given by Eq. (\ref{pr22}). It is clear that this asymptotic value corresponds to the steady state of the system under a weak external potential $\hat\Phi_e({\bf k})$\footnote{It suffices to take $\partial/\partial t=0$ in the linearized equations of Sections \ref{sec_lde} and \ref{sec_vl}. The equations of this section remain valid provided that we use just Fourier transforms and set  $\omega=0$. Eqs. (\ref{pr14mar}) and (\ref{vl3bma}) then yield Eq.  (\ref{pr22}).}. We can make this expression more explicit for a collision-dominated gas and a collisionless system.

\subsection{Collision-dominated gas}
\label{sec_asycoll}

For a collision-dominated gas, the dielectric function (\ref{lde9}) with $\omega=0$ reduces to
\begin{equation}
\label{asy1}
\epsilon({k},0)=1+\frac{(2\pi)^d\hat{u}(k)\rho}{c_s^2}.
\end{equation}
Substituting this expression in Eq. (\ref{pr22}), we get
\begin{equation}
\label{asy2}
\delta\hat{\Phi}_{\infty}({\bf k})=-\frac{(2\pi)^d\hat{u}(k)\rho k^2}{c_s^2k^2+(2\pi)^d\hat{u}(k)k^2\rho}\hat{\Phi}_e({\bf k}).
\end{equation}
We can show that Eq. (\ref{asy2}) corresponds to the steady state of the damped Euler equations (\ref{de1})-(\ref{de3}) with the equation of state $p(\rho)$ under a weak external field $\Phi_e({\bf r})$. Indeed, in the presence of an external field, the condition of hydrostatic equilibrium reads
\begin{equation}
\label{asy3}
\nabla p+\rho\nabla\Phi+\rho\nabla\Phi_e={\bf 0}.
\end{equation}
If the applied field is weak, we can linearize Eq. (\ref{asy3}) about the unperturbed distribution. If the unperturbed system is spatially homogeneous, we obtain
\begin{equation}
\label{asy3b}
c_s^2\nabla \delta\rho+\rho\nabla\delta\Phi+\rho\nabla\Phi_e={\bf 0}.
\end{equation}
After integration, we get
\begin{equation}
\label{asy4}
c_s^2\delta\rho({\bf r})+\rho(\delta\Phi({\bf r})-\langle \delta\Phi\rangle)+\rho (\Phi_e({\bf r})-\langle \Phi_e\rangle)=0,
\end{equation}
where the constant of integration has been determined by using the fact that the external potential does not change the mass (the brackets denote a space  average over the whole domain).  Taking the Fourier transform of this expression, and using Eq. (\ref{lde6}), we get Eq. (\ref{asy2}). Therefore, when submitted to a step function, a stable collision-dominated gas converges for $t\rightarrow +\infty$ towards an asymptotic distribution $\delta\hat{\Phi}_{\infty}({\bf k})$ which is the steady state of the damped Euler equations (\ref{de1})-(\ref{de3}) under a weak external field $\Phi_e({\bf r})$\footnote{This is true for the damped Euler equation ($\xi>0$) since all the zeros of the dielectric function lie in the lower half-plane. For the Euler equation ($\xi=0$), the perturbation oscillates indefinitely about $\delta\hat{\Phi}_{\infty}({\bf k})$ as explained in Section \ref{sec_mar}.}.

\subsection{Collisionless system}
\label{sec_asyless}

For a collisionless system, the dielectric function (\ref{vl6}) with $\omega=0$ reduces to
\begin{equation}
\label{asy5}
\epsilon({k},0)=1-(2\pi)^d\hat{u}(k)\int \frac{f'(v)}{v}\, dv.
\end{equation}
Substituting this expression in Eq. (\ref{pr22}), we get
\begin{equation}
\label{asy6}
\delta\hat{\Phi}_{\infty}({\bf k})=\frac{(2\pi)^d\hat{u}(k)\int \frac{f'(v)}{v}\, dv}{1-(2\pi)^d\hat{u}(k)\int \frac{f'(v)}{v}\, dv}\hat{\Phi}_e({\bf k}).
\end{equation}
When $f({\bf v})=F(v^2/2)$, we can show that Eq. (\ref{asy6}) corresponds to the steady state of the Vlasov equation (\ref{v1})-(\ref{v2}) with the distribution $F$ under a weak external field $\Phi_e({\bf r})$. Indeed, from the spatially homogeneous distribution function $f({\bf v})=F(v^2/2)$, we can define a spatially inhomogeneous distribution function $f({\bf r},{\bf v})=F(\epsilon)$ with $\epsilon=v^2/2+\Phi_t({\bf r})$ where $\Phi_t({\bf r})=\Phi({\bf r})+\Phi_e({\bf r})$ is the total potential. This distribution $f({\bf r},{\bf v})=F(\epsilon)$ is a steady state of the Vlasov equation. We have to show that, when the external potential is weak, this distribution function leads to Eq. (\ref{asy6}). To that purpose, we first note that, for any collisionless system with a distribution function of the form $f=F(\epsilon)$, there exist a corresponding barotropic gas $p=p(\rho)$ with the same density profile (see Appendix \ref{sec_corr}).  Furthermore, the condition $f=F(\epsilon)$ implies the condition of  hydrostatic equilibrium (\ref{asy3}). Therefore, when the external potential is weak, proceeding as in Section \ref{sec_asycoll}, we find that the perturbation is given by
\begin{equation}
\label{asy7}
\delta\hat{\Phi}_{\infty}({\bf k})=-\frac{(2\pi)^d\hat{u}(k)\rho k^2}{c_s^2k^2+(2\pi)^d\hat{u}(k)k^2\rho}\hat{\Phi}_e({\bf k}).
\end{equation}
Now, the velocity of sound in the corresponding  barotropic gas is given by Eq. (\ref{corr3}). Substituting this relation in Eq. (\ref{asy7}), we recover Eq. (\ref{asy6})\footnote{If we expand $F(\epsilon)=F(v^2/2+\Phi_t({\bf r}))$ for $\Phi_t({\bf r})\ll 1$, we obtain  Eq. (\ref{vl3b}) with $\omega=0$. After integration over ${\bf v}$, this leads to Eq. (\ref{vl3bma}) which is equivalent to Eq. (\ref{asy6}). This shows the compatibility of the different approaches.}. Therefore, when submitted to a step function, a stable collisionless system converges for $t\rightarrow +\infty$  towards an asymptotic distribution $\delta\hat{\Phi}_{\infty}({k})$ which is the steady state of the Vlasov equation  (\ref{v1})-(\ref{v2}) with the distribution $F$ under a weak external field $\Phi_e({\bf r})$\footnote{This is true for most distribution functions (that experience Landau damping). A notorious exception is the waterbag distribution that does not experience Landau damping \cite{patelli}. In that case, the perturbation oscillates indefinitely about $\delta\hat{\Phi}_{\infty}({\bf k})$ as in a perfect gas ($\xi=0$).}. Using Eq. (\ref{corr3}), we note that the dielectric function (\ref{asy5}) can be written as
\begin{equation}
\label{asy8}
\epsilon({k},0)=1+\frac{(2\pi)^d\hat{u}(k)\rho}{c_s^2},
\end{equation}
just like  in a collision-dominated gas.

\subsection{The relation with the correlation function}

For the isothermal distribution function (\ref{app1}), corresponding to statistical equilibrium, the asymptotic expression of the perturbation reduces to
\begin{equation}
\label{asy9}
\delta\hat{\Phi}_{\infty}({\bf k})=-\frac{(2\pi)^d\hat{u}(k)\rho k^2}{T k^2+(2\pi)^d\hat{u}(k)k^2\rho}\hat{\Phi}_e({\bf k}).
\end{equation}
As a confirmation of the previous calculations, we can directly establish  that $\delta\hat{\Phi}_{\infty}({\bf k})$ corresponds to the potential produced by the Boltzmann distribution with a weak external field. For a system at statistical equilibrium under an external field, the Boltzmann distribution function reads
\begin{equation}
\label{asy10}
f({\bf r},{\bf v})=Ae^{-\beta \lbrack v^2/2+\Phi({\bf r})+\Phi_e({\bf r})\rbrack}.
\end{equation}
Integrating over the velocity, we get
\begin{equation}
\label{asy11}
\rho({\bf r})=N\frac{e^{-\beta \lbrack\Phi({\bf r})+\Phi_e({\bf r})\rbrack}}{\int e^{-\beta \lbrack\Phi({\bf r})+\Phi_e({\bf r})\rbrack}\, d{\bf r}}.
\end{equation}
If the external field is weak, we can linearize the previous equation. Assuming that the unperturbed system is spatially homogeneous, we obtain
\begin{equation}
\label{asy12}
\delta\rho({\bf r})=-\beta\rho (\delta\Phi({\bf r})-\langle \delta\Phi\rangle)-\beta\rho (\Phi_e({\bf r})-\langle \Phi_e\rangle).
\end{equation}
This returns Eq. (\ref{asy3b}) with $c_s^2=T$, leading to Eq. (\ref{asy9}). Therefore, Eq. (\ref{asy9}) corresponds to the statistical equilibrium state of the system under a weak external potential. We stress, however, that Eq. (\ref{asy6}) [resp. Eq. (\ref{asy7})] is more general since it is valid for an arbitrary distribution $f(v)$ [resp. $f({\bf v})=F(v^2/2)$], not only for the Boltzmann distribution.

Finally, if we recall the expression of the Fourier transform of the two-body correlation function of a spatially homogeneous system with long-range interactions \cite{hb1,paper5}:
\begin{equation}
\label{asy9bb}
(2\pi)^d n \hat{h}({k})=-\frac{(2\pi)^d \rho \hat{u}(k)k^2}{T k^2+(2\pi)^d\hat{u}(k)k^2\rho}=\frac{1-\epsilon({k},0)}{\epsilon({k},0)},
\end{equation}
and compare its expression with Eq. (\ref{asy9}), we obtain the relation
\begin{equation}
\label{asy9cc}
\delta\hat{\Phi}_{\infty}({\bf k})=(2\pi)^d\hat{h}(k) n \hat{\Phi}_e({\bf k}),
\end{equation}
where $n=\rho/m$ in the numerical density. Therefore, the measure of the asymptotic potential produced by  a system submitted to a step function allows us to determine the equilibrium correlation function of the system.

\section{Application to specific distribution functions}
\label{sec_app}

In the previous sections, we have given general criteria of stability and general expressions for the asymptotic distribution of a stable homogeneous system submitted to step function. These results are valid for arbitrary distributions and for arbitrary potentials of interaction. Furthermore, when $f({\bf v})=F(v^2/2)$, we have shown that the results take the same form in a collisionless system and in the corresponding barotropic gas. In particular, they can be expressed very simply in terms of the velocity of sound [see Eqs. (\ref{drv3}) and (\ref{asy7})]. As explained in Appendix \ref{sec_corr}, the velocity of sound is a function of a (generalized) temperature $T$. In turn, the temperature  can be expressed in terms of the energy $E$ by Eq. (\ref{corr4}). In this section, we illustrate these results for specific distribution functions (isothermal, polytropic, waterbag, and Fermi-Dirac). In the context of the HMF model, we can derive simple formulae that complete those obtained previously in Refs. \cite{patelli,oyresponse}. In particular, we obtain the general result
\begin{equation}
\label{gen1}
M_{\infty}=\frac{h}{2c_s^2-1}, \qquad  c_s^2>\frac{1}{2},
\end{equation}
for the asymptotic magnetization\footnote{For the HMF model, it is more convenient to work in terms of the magnetization $M$ than in terms of the potential $\delta\Phi(\theta)$. They are related to each other by $\delta\Phi(\theta)=-M\cos\theta$ yielding $\delta\hat\Phi_n=-({M}/{2})\delta_{n,\pm 1}$. It is also convenient to write the external potential in the form $\Phi_e=-h\cos\theta$, yielding  $(\hat\Phi_{e})_n=-(h/2)\delta_{n,\pm 1}$, where $h$ can be interpreted as a ``magnetic'' field.}. The asymptotic magnetization $M_{\infty}$ is proportional to the magnetic field $h$, and the magnetic susceptibility $\chi=M_{\infty}/h$ is given by a generalized Curie-Weiss law: $\chi=1/(2 c_s^2-1)$.

\subsection{Isothermal distribution}
\label{sec_appiso}

For the isothermal distribution function
\begin{equation}
\label{app1}
f({\bf v})=\left (\frac{\beta}{2\pi}\right )^{d/2}\rho\, e^{-\frac{1}{2}\beta v^2},
\end{equation}
we obtain the equation of state
\begin{equation}
\label{app2}
p=\rho T,
\end{equation}
and the velocity of sound
\begin{equation}
\label{app3}
c_s^2=T.
\end{equation}
For the HMF model, we can substitute this result in Eq. (\ref{gen1}) to express the stability criterion and the asymptotic magnetization in terms of the temperature $T$. We obtain
\begin{equation}
\label{app4}
M_{\infty}=\frac{h}{2(T-T_c)}, \qquad T>T_c=\frac{1}{2},
\end{equation}
\begin{equation}
\label{app5}
M_{\infty}=\frac{h}{4(E-E_c)}, \qquad E>E_c=\frac{3}{4},
\end{equation}
where we have used $E=T/2+1/2$ [see Eq. (\ref{corr4})]. This returns the results obtained in \cite{patelli,oyresponse}. Eqs. (\ref{app4})-(\ref{app5}) are also identical to the expressions of the magnetization of the HMF model at statistical equilibrium under a {\it weak} magnetic field $h$ (see Eqs. (100) and (107) of \cite{hmfmagnetique}). As we have explained in Section \ref{sec_asy}, this property is general.

\subsection{Polytropic distributions}
\label{sec_poly}

The polytropic distributions can be written as \cite{cc}:
\begin{eqnarray}
f={1\over Z}\biggl \lbrack \rho^{1/n}-{v^{2}/2\over (n+1)K}\biggr\rbrack_+^{n-d/2},
\label{app6}
\end{eqnarray}
where $Z$ is given for $n\ge d/2$ by
\begin{eqnarray}
Z=S_d 2^{d/2-1}{\Gamma(d/2)\Gamma(1-d/2+n)\over\Gamma(n+1)}\lbrack K(n+1)\rbrack^{d/2},
\label{app7}
\end{eqnarray}
and for $n<-1$ by
\begin{eqnarray}
Z=S_d 2^{d/2-1}{\Gamma(-n)\Gamma(d/2)\over\Gamma(d/2-n)}\lbrack -K(n+1)\rbrack^{d/2}.
\label{app8}
\end{eqnarray}
The constant $K$ is called the polytropic temperature. These
distribution functions were also introduced by Tsallis \cite{tsallis}
in his generalized thermodynamics. The corresponding equation of state
is
\begin{eqnarray}
p=K\rho^{\gamma},\qquad \gamma=1+\frac{1}{n}.
\label{app9}
\end{eqnarray}
The velocity of sound is given by
\begin{equation}
\label{app10}
c_s^2=K\gamma\rho^{\gamma-1}.
\end{equation}
For the HMF model, we can substitute this result in Eq. (\ref{gen1}) to express the stability criterion and the asymptotic magnetization in terms of the polytropic temperature $K$. Since $\rho=1/2\pi$ in the homogeneous phase, it is convenient to define the polytropic temperature by $\Theta=K(1/2\pi)^{\gamma-1}$ so that $c_s^2=\gamma\Theta$. Then, we  obtain
\begin{equation}
\label{app11}
M_{\infty}=\frac{h}{2\gamma\Theta-1},\qquad  \Theta>\frac{1}{2\gamma},
\end{equation}
\begin{equation}
\label{app12}
M_{\infty}=\frac{h}{4\gamma E-2\gamma-1},\qquad  E>\frac{1}{4\gamma}+\frac{1}{2},
\end{equation}
where we have used $E=\Theta/2+1/2$ [see Eq. (\ref{corr4})]. For $n\rightarrow +\infty$, the distribution function (\ref{app6}) reduces to the isothermal distribution (\ref{app1}) and we recover the results of Section \ref{sec_appiso}. 

\subsection{Waterbag distribution}
\label{sec_water}

The waterbag distribution, defined by $f=\eta_0$ for $-v_m<v<v_m$ and $f=0$ otherwise, is a particular polytrope of index $n=d/2$. The corresponding density and pressure are $\rho=\eta_0 S_d v_m^d/d$ and $p=\eta_0 S_d v_m^{d+2}/\lbrack d(d+2)\rbrack$, leading to the polytropic equation of state
\begin{equation}
\label{water0}
p=K\rho^{\gamma},\qquad \gamma=\frac{d+2}{d}, \qquad K=\frac{1}{d+2}\left (\frac{d}{\eta_0 S_d}\right )^{2/d}.
\end{equation}
The velocity of sound is $c_s^2=v_m^2/d$. In $d=1$, we have $c_s=v_m$
in agreement with the results of Section \ref{sec_waterbag}. For the
HMF model, we can substitute these results in the general equation
(\ref{gen1}) to express the stability criterion and
the asymptotic magnetization in terms of $v_m$. We
obtain
\begin{equation}
\label{water1}
M_{\infty}=\frac{h}{2v_m^2-1},\qquad  v_m>\frac{1}{\sqrt{2}},
\end{equation}
\begin{equation}
\label{water2}
M_{\infty}=\frac{h}{12 E-7},\qquad  E>\frac{7}{12},
\end{equation}
where we have used $E=v_m^2/6+1/2$ [see Eq. (\ref{corr4})]. We recall, however, that a collisionless system with the waterbag distribution submitted to a step function does {\it not} relax towards the magnetization $M_{\infty}$ but oscillates about it as $M(t)=M_{\infty}\lbrack 1-\cos(\omega_0 t)\rbrack$ with the pulsation $\omega_0=\pm \sqrt{v_m^2-1/2}$ (see Eq. (\ref{se4}) with the remark of Section \ref{sec_waterbag}). This result was obtained in \cite{patelli} and it was confirmed by direct numerical simulations. When submitted to a pulse, the magnetization behaves as $M(t)=-\sin(\omega_0 t)/\omega_0$ according to Eq. (\ref{pe5}).

\subsection{Fermi-Dirac and Lynden-Bell distributions}

For the Fermi-Dirac distribution function (see, e.g., \cite{fermid}): 
\begin{equation}
\label{app13}
f({\bf v})=\frac{\eta_0}{1+\lambda e^{\frac{1}{2}\beta v^2}},
\end{equation}
we obtain
\begin{equation}
\label{app14}
\rho=\frac{\eta_0 S_d 2^{d/2-1}}{\beta^{d/2}}I_{d/2-1}(\lambda),\qquad p=\frac{\eta_0 S_d 2^{d/2}}{d\beta^{d/2+1}}I_{d/2}(\lambda),
\end{equation}
where $\eta_0$ is the maximum value of the distribution function fixed by the Pauli exclusion principle, $S_d$ is the surface of a unit sphere in $d$ dimensions, and $I_n(\lambda)=\int_0^{+\infty}x^n/(1+\lambda e^x)\, dx$ are the Fermi integrals. This distribution function also corresponds to the prediction of Lynden-Bell \cite{lb} in his statistical theory of the violent relaxation of the Vlasov equation. In that context, $\eta_0$ represents the initial value of the distribution function (in the two-levels case). Equations (\ref{app13}) and (\ref{app14}) define the equation of state $p(\rho)$ of the Fermi gas in parametric form. In the non degenerate limit ($T\rightarrow +\infty$), the Fermi-Dirac distribution function reduces to the classical isothermal distribution function (\ref{app1}). The corresponding equation of state is $p=\rho T$ (see Section \ref{sec_appiso}). In the completely degenerate limit ($T\rightarrow 0$), the Fermi-Dirac distribution function reduces to the waterbag distribution $f=\eta_0$ for $-v_F<v<v_F$ and $f=0$ otherwise, where $v_F$ is the Fermi velocity (see Section \ref{sec_water}). The corresponding equation of state is the polytropic equation of state (\ref{water0}) with $v_m=v_F$. The velocity of sound can be written as
\begin{equation}
\label{app15}
c_s^2=\frac{2}{d\beta}\frac{I_{d/2}'(\lambda)}{I_{d/2-1}'(\lambda)}.
\end{equation}
For a given temperature $T$, we can determine $\lambda$ from Eq. (\ref{app14}-a) and $c_s$ from Eq. (\ref{app15}). Therefore, the velocity of sound is a function of the temperature: $c_s=c_s(T)$. In the non degenerate limit, $c_s^2=T$ and in the completely degenerate limit $c_s^2=v_F^2/d$. For the HMF model, we can substitute this result in Eq. (\ref{gen1}) to express the stability criterion and the asymptotic magnetization in terms of the temperature $T$. We obtain
\begin{equation}
\label{app16}
M_{\infty}=\frac{h}{2c_s^2(T)-1}, \qquad  c_s^2(T)>\frac{1}{2}.
\end{equation}
Using Eqs. (\ref{corr4}) and (\ref{app14}-b), we find that the energy is given by
\begin{equation}
\label{app17}
E=\frac{\pi\eta_0 2\sqrt{2}}{\beta^{3/2}}I_{1/2}(\lambda)+\frac{1}{2}.
\end{equation}
Eliminating $\lambda$ between Eqs. (\ref{app17}) and (\ref{app14}-a), we can obtain $E(T)$ in parametric form. We can then express the results of Eq. (\ref{app16}) in terms of $E$ instead of $T$. In the Lynden-Bell theory, $\eta_0$ is an additional control parameter related to the initial condition, so that the results actually depend on $(T,\eta_0)$ or on $(E,\eta_0)$. The stability of the spatially homogeneous distribution [see criterion (\ref{app16}-b)] was studied in \cite{epjblb}. For $T=0$, corresponding to $E=E_{min}=1/(96\pi^2\eta_0^2)+1/2$, we find that the system is stable for $\eta_0<1/(2\sqrt{2}\pi)$. In that case, $M_{\infty}=h/[1/(8\pi^2\eta_0^2)-1]$ (this corresponds to the results of Section \ref{sec_water} with $v_m=v_F=1/(4\pi\eta_0)$). The asymptotic expression of the magnetization $M_{\infty}(T)$ in the limit $T\rightarrow 0$ or $E\rightarrow E_{min}$ (for fixed $\eta_0$) was obtained in \cite{patelli}. Using the Sommerfeld expansions of the Fermi integrals (see, e.g., Eqs. (39) and (40) of \cite{epjblb}), we can check that Eq. (\ref{app16}) returns the results of \cite{patelli}.  However, Eq. (\ref{app16}) is more general, as it is valid at any temperature $T\ge 0$ or at any energy $E\ge E_{min}$.

\section{The initial value problem}
\label{sec_iv}

In Section \ref{sec_vlasov}, we have studied the response of a collisionless system described by the Vlasov equation to a weak external potential $\Phi_e({\bf r},t)$ using the linear response theory. Here, we compare these results with those obtained when the system is isolated (i.e., $\Phi_e({\bf r},t)=0$), but the distribution function is slightly perturbed at $t=0$. This is the so-called initial value problem of the linearized Vlasov equation \cite{balescubook}. 

Taking the Fourier-Laplace transform of the linearized Vlasov equation (\ref{vl1})-(\ref{vl2}), and assuming now that $\delta f({\bf r},{\bf v},0)\neq 0$, we obtain
\begin{equation}
\label{iv1}
\delta\tilde{f}({\bf k},{\bf v},\omega)=\frac{{\bf k}\cdot \frac{\partial f}{\partial {\bf v}}}{{\bf k}\cdot {\bf v}-\omega} \delta\tilde\Phi({\bf k},\omega)+\frac{\delta\hat{f}({\bf k},{\bf v},0)}{i({\bf k}\cdot {\bf v}-\omega)},
\end{equation}
where $\delta\hat{f}({\bf k},{\bf v},0)$ is the Fourier transform of the initial perturbation $\delta f({\bf r},{\bf v},0)$. Integrating Eq. (\ref{iv1}) over the velocity and using Eq. (\ref{vl4}), we find that
\begin{equation}
\label{iv2}
\delta\tilde{\Phi}({\bf k},\omega)=(2\pi)^d\frac{\hat{u}(k)}{\epsilon({\bf k},\omega)}\int \frac{\delta\hat{f}({\bf k},{\bf v},0)}{i({\bf k}\cdot {\bf v}-\omega)}\, d{\bf v}.
\end{equation}
Substituting this result back  into Eq. (\ref{iv1}), we get
\begin{equation}
\label{iv3}
\delta\tilde{f}({\bf k},{\bf v},\omega)=\frac{{\bf k}\cdot \frac{\partial f}{\partial {\bf v}}}{{\bf k}\cdot {\bf v}-\omega}(2\pi)^d\frac{\hat{u}(k)}{\epsilon({\bf k},\omega)}\int \frac{\delta\hat{f}({\bf k},{\bf v}',0)}{i({\bf k}\cdot {\bf v}'-\omega)}\, d{\bf v}'+\frac{\delta\hat{f}({\bf k},{\bf v},0)}{i({\bf k}\cdot {\bf v}-\omega)}.
\end{equation}
This is the exact solution of the initial value problem for the linearized Vlasov equation. It is conveniently written in terms of a resolvent operator that connects $\delta\tilde{f}({\bf k},{\bf v},\omega)$ to the initial value
\begin{equation}
\label{iv4}
\delta\tilde{f}({\bf k},{\bf v},\omega)=\int d{\bf v}' R({\bf v}|{\bf v}',{\bf k},\omega) \delta\hat{f}({\bf k},{\bf v},0),
\end{equation}
with
\begin{equation}
\label{iv5}
R({\bf v}|{\bf v}',{\bf k},\omega)=\frac{{\bf k}\cdot \frac{\partial f}{\partial {\bf v}}}{{\bf k}\cdot {\bf v}-\omega}(2\pi)^d\frac{\hat{u}(k)}{\epsilon({\bf k},\omega)}\frac{1}{i({\bf k}\cdot {\bf v}'-\omega)}+\frac{\delta({\bf v}-{\bf v}')}{i({\bf k}\cdot {\bf v}-\omega)}.
\end{equation}
If we consider an initial condition of the form $\delta f({\bf r},{\bf v},0)=m\delta({\bf v}-{\bf v}')\delta({\bf r}-{\bf r}')$, leading to
\begin{equation}
\label{iv6}
\delta\hat{f}({\bf k},{\bf v},0)=\frac{m}{(2\pi)^d}\delta({\bf v}-{\bf v}')e^{-i{\bf k}\cdot {\bf r}'},
\end{equation}
we find that
\begin{equation}
\label{iv7}
\delta\tilde{f}({\bf k},{\bf v},\omega)=\frac{{\bf k}\cdot \frac{\partial f}{\partial {\bf v}}}{{\bf k}\cdot {\bf v}-\omega}\frac{\hat{u}(k)}{\epsilon({\bf k},\omega)}m \frac{e^{-i{\bf k}\cdot {\bf r}'}}{i({\bf k}\cdot {\bf v}'-\omega)}+\frac{m}{(2\pi)^d}\delta({\bf v}-{\bf v}')\frac{e^{-i{\bf k}\cdot {\bf r}'}}{i({\bf k}\cdot {\bf v}-\omega)}.
\end{equation}
On the other hand, if we substitute Eq. (\ref{vl4}) in Eq. (\ref{iv1}), we obtain
\begin{equation}
\label{iv8}
\delta\tilde{f}({\bf k},{\bf v},\omega)=\frac{{\bf k}\cdot \frac{\partial f}{\partial {\bf v}}}{{\bf k}\cdot {\bf v}-\omega} (2\pi)^d\hat{u}(k)\int \delta\tilde{f}({\bf k},{\bf v}',\omega)\, d{\bf v}'+\frac{\delta\hat{f}({\bf k},{\bf v},0)}{i({\bf k}\cdot {\bf v}-\omega)}.
\end{equation}
This is an integral equation for $\delta\tilde{f}({\bf k},{\bf v},\omega)$ that is equivalent to Eq. (\ref{iv3}).

These results are well-known in plasma physics
\cite{balescubook}. They have been recalled in order to facilitate the
comparison with the results obtained with the linear response
theory. In particular, Eq. (\ref{iv2}) may be compared to
Eq. (\ref{vl3bma}) and Eq. (\ref{iv3}) may be compared to
Eq. (\ref{vl3b}). In general, the two sets of equations differ,
showing that the initial value problem is not equivalent to the linear
response theory. However, if we consider an initial disturbance of the
form
\begin{equation}
\label{iv9}
\delta\hat{f}({\bf k},{\bf v},0)=i{\bf k}\cdot \frac{\partial f}{\partial {\bf v}}\hat\Phi_e({\bf k}),
\end{equation}
where $\hat\Phi_e({\bf k})$ is an arbitrary function of ${\bf k}$ (independent on ${\bf v}$), we see that Eqs. (\ref{vl3}) and (\ref{iv1}) coincide provided that $\tilde\Phi_e({\bf k},\omega)=\hat\Phi_e({\bf k})$. Since the Laplace transform of the external potential is independent on $\omega$, it corresponds to a pulse with an amplitude $\hat\Phi_e({\bf k})$ (see Section \ref{sec_pulss}). Therefore, the effect of an initial disturbance of the form (\ref{iv9}) on an isolated system is equivalent to submitting this system to a pulse $\hat\Phi_e({\bf k},t)=\hat\Phi_e({\bf k})\delta(t)$.

The perturbation $\delta\tilde{\Phi}({\bf k},\omega)$ given  by Eq. (\ref{iv2}) appears as a product of two factors: $\epsilon({\bf k},\omega)^{-1}$ and an integral involving the initial condition $\delta\hat{f}({\bf k},{\bf v},0)$. The first factor is due to collective effects, and its role in the evolution of the perturbation has been already discussed in Section \ref{sec_pulss}. This term can produce damped, steady, or growing oscillations. On the other hand, the integral corresponds to the excess density produced by an initial disturbance in a gas of non-interacting particles (i.e., for which $\epsilon({\bf k},\omega)=1$). It is typical of an individual particle behavior. We expect that the effect of this term will disappear for late times. Actually, it can be shown that the integral produces damped oscillations (i.e. its poles are necessarily in the lower half plane or on the real axis) \cite{balescubook}. This is associated with the phenomenon of ``phase mixing'' which is an irreversible homogenization process in which the interactions play no role\footnote{If we consider the perturbed distribution function (\ref{iv3}), we see that there is a real pole $\omega={\bf k}\cdot {\bf v}$. It produces an undamped oscillation ${\rm exp}(-i {\bf k}\cdot {\bf  v} t)$ whose pulsation is proportional to the velocity ${\bf v}$ of the particles. It is therefore an individual particle effect. Let us consider the excess density $\delta\tilde\rho({\bf k},t)=\int \delta\tilde f({\bf k},{\bf v},t)\, d{\bf v}$. The contribution to this integral from the various velocities will produce destructive interferences of the oscillations, hence this part of $\delta\tilde\rho({\bf k},t)$ tends to zero for long times: this is the phenomenon of phase mixing. The other poles of Eq. (\ref{iv3}) correspond to the zeros of the dielectric function: $\epsilon({\bf k},\omega)=0$. They depend on ${\bf k}$ but not on ${\bf v}$. They describe the collective behavior of the system. They produced damped or growing oscillations that ``resist'' the integration over ${\bf v}$ \cite{balescubook}.}. In a sense, this term is the counterpart of the polarization function in the linear response theory.

The individual particle behavior corresponds to the integral in Eq. (\ref{iv2}). This integral represents the excess density in the absence of interaction. It can be written as
\begin{equation}
\label{iv10}
\delta\tilde{\rho}_0(k,\omega)=\int_L \frac{\delta\hat{f}(k,v,0)}{i(k v-\omega)}\, dv,
\end{equation}
where $L$ denotes the Landau contour and $\delta\hat{f}(k,v,0)$ is the reduced distribution function corresponding to the initial disturbance (i.e. it has been integrated over the coordinates perpendicular to ${\bf k}\cdot {\bf v}$).
The individual particle behavior can be studied analytically for an initial disturbance of the form
\begin{equation}
\label{iv11}
\delta\hat{f}(k,v,0)=\frac{A(k)}{\pi w_0}\frac{1}{1+\frac{(v-v_0)^2}{w_0^2}}.
\end{equation}
This example has been discussed in Ref. \cite{balescubook}. Using the residue theorem, we easily obtain
\begin{equation}
\label{iv12}
\delta\tilde{\rho}_0(k,\omega)=\frac{A(k)}{i(kv_0-ikw_0-\omega)}.
\end{equation}
Then, taking the inverse Laplace transform of this expression, we get
\begin{equation}
\label{iv13}
\delta\hat{\rho}_0(k,t)=A(k)e^{-ikv_0 t}e^{-kw_0t}.
\end{equation}
This corresponds to a damped oscillation: The perturbation propagates with a group velocity $v_0$ and dies out exponentially rapidly on a timescale $\tau_k=(k w_0)^{-1}$ depending on the wavenumber $k$. We now treat another example that has not been discussed before (to the best of our knowledge). We consider an initial disturbance of the form $\delta\hat{f}(k,v,0)=A(k)/2w_m$ if $-w_m\le v\le w_m$ and $\delta\hat{f}(k,v,0)=0$ otherwise. The excess density (\ref{iv10}) can be written explicitly as
\begin{equation}
\label{iv14}
\delta\tilde{\rho}_0(k,\omega)=\frac{A(k)}{2w_m ik}\ln\left (\frac{\omega-kw_m}{\omega+kw_m}\right ).
\end{equation}
Taking the inverse Laplace transform of this expression, and integrating by parts, we obtain
\begin{equation}
\label{iv15}
\delta\hat{\rho}_0(k,t)=A(k)\frac{\sin(k w_m t)}{k w_m t}.
\end{equation}
We see that the damping of the initial perturbation is {\it
algebraic}, behaving as $t^{-1}$ for large times, instead of being
exponential as in Eq. (\ref{iv13}). It also exhibits
oscillations with a pulsation $k w_m$ that depends on the wavenumber
$k$. As a result of this slow damping, the initial disturbance will
have a long-term effect.

It is instructive to recover these results in a different manner \cite{balescubook}. Taking the Laplace transform of Eq. (\ref{iv10}) before performing the velocity integration, we get
\begin{equation}
\label{iv16}
\delta\hat{\rho}_0(k,t)=\int \delta\hat{f}(k,v,0)e^{-i k v t}\, dv.
\end{equation}
Under this form, we see that the evolution corresponding to the free-particle motion consists of a superposition of waves ${\rm exp}(-i k v t)$ propagating with a group velocity equal to the velocity of the individual particles. These waves will in general interfere destructively as time goes on, leading to the free-motion damping discussed above. Actually, the integral (\ref{iv16}) can be calculated easily returning the results (\ref{iv13}) and (\ref{iv15}).

Finally, as shown in \cite{balescubook}, when the unperturbed distribution function is the Cauchy distribution (\ref{c1}), and when the initial disturbance in given by Eq. (\ref{iv11}), it is possible to determine the evolution of the perturbation $\delta\hat{\Phi}(k,t)$ given by Eq. (\ref{iv2}) analytically. Generalizing the calculation to an arbitrary potential, we get
\begin{eqnarray}
\label{iv17}
\delta\hat{\Phi}(k,t)=(2\pi)^d\hat{u}(k)A(k)\biggl\lbrace e^{-ikv_0t-kw_0t}\left\lbrack 1+\frac{\Omega^2}{(k v_0+i(u_0-w_0)k)^2-\Omega^2}\right\rbrack\nonumber\\
+\frac{1}{2}\Omega e^{-k u_0 t}\left\lbrack \frac{-e^{-i\Omega t}}{k v_0+i k(u_0-w_0)-\Omega}+\frac{e^{i\Omega t}}{k v_0+i k(u_0-w_0)+\Omega}\right\rbrack\biggr\rbrace,
\end{eqnarray}
where we have defined the pulsation $\Omega(k)=\sqrt{(2\pi)^d \hat{u}(k)k^2\rho}$ which is real for a repulsive interaction ($\hat{u}(k)>0$) and purely imaginary for an attractive interaction ($\hat{u}(k)<0$).   The first term in braces in Eq. (\ref{iv17})  corresponds to individual effects and the second term corresponds to collective effects.

\section{Conclusion}

We have applied the linear response theory to systems with long-range
interactions. Our study completes previous investigations 
\cite{patelli,oyresponse}. We have considered a collision-dominated
gas (Euler) and a collisionless system (Vlasov). We have shown that
the response of these systems to an external field is in general
different except for the waterbag distribution. In that case, there is
no Landau damping and the dielectric function of the waterbag
distribution coincides with the dielectric function of a
collision-dominated gas without dissipation ($\xi=0$). When submitted
to a step function, these systems oscillate permanently without
reaching a steady state. For more generic distributions, such as the
Cauchy distribution, there is Landau damping and the evolution of a
collisionless system described by the Vlasov equation resembles the
evolution of a collision-dominated gas described by the damped Euler
equation ($\xi>0$). When submitted to a step function, these systems
relax towards a steady distribution. When $f=F(v^2/2)$ this
distribution is the same in the collisionless system and in the
corresponding barotropic gas. It corresponds to the steady state of
the system under a weak external field. However, the relaxation
towards this steady state is in general different in a collisional gas
and in a collisionless gas because their dielectric functions
differ. We have also considered unstable systems. In that case, the
linear response theory is valid only for short times, before the
perturbation has significatively grown.  Physical applications of the
linear response theory will be given in a future paper \cite{prep}.

\appendix

\section{The corresponding barotropic gas}
\label{sec_corr}

We consider a collisionless system described by the Vlasov equation (\ref{v1})-(\ref{v2}). We assume that the system is spatially homogeneous. Any distribution function $f({\bf v})$ is a steady state of the Vlasov equation. We restrict ourselves to distribution functions of the form $f({\bf v})=F(v^2/2)$.  From the spatially homogeneous distribution function $f({\bf v})=F(v^2/2)$, we can define a spatially inhomogeneous distribution function $f({\bf r},{\bf v})=F(\epsilon)$ where $\epsilon=v^2/2+\Phi({\bf r})$ is the individual energy of the particles.  This distribution $f({\bf r},{\bf v})=F(\epsilon)$ is also a steady state of the Vlasov equation.

For any collisionless system with a distribution function of the form $f({\bf r},{\bf v})=F(\epsilon)$, there exist a corresponding barotropic gas with the same equilibrium density \cite{lbsanitt}. Indeed, introducing the density $\rho({\bf r})=\int F(\epsilon)\, d{\bf v}$ and the pressure $p({\bf r})=\frac{1}{d}\int F(\epsilon) v^2\, d{\bf v}$, we have $\rho({\bf r})=\rho\lbrack \Phi({\bf r})\rbrack$ and $p({\bf r})=p\lbrack \Phi({\bf r})\rbrack$. Eliminating the potential $\Phi({\bf r})$ between these two expressions, we obtain a barotropic equation of state $p=p(\rho)$ that is entirely determined by the function $F$. We can easily show that the condition $f({\bf r},{\bf v})=F(\epsilon)$ implies the condition of  hydrostatic equilibrium. Indeed
\begin{equation}
\label{corr1}
\nabla p=\frac{1}{d}\nabla\Phi\int F'(\epsilon)v^2\, d{\bf v}=\frac{1}{d}\nabla\Phi\int \frac{\partial f}{\partial {\bf v}}\cdot {\bf v}\, d{\bf v}=-\nabla\Phi\int f\, d{\bf v}=-\rho \nabla\Phi.
\end{equation}
Finally, we can relate the velocity of sound $c_s^2=p'(\rho)$ in the corresponding barotropic gas to the distribution function $f({\bf v})$. To that purpose, we first note that the condition of hydrostatic equilibrium (\ref{corr1}) combined with the equation of state $p=p(\rho)$ implies that $p'(\Phi)=-\rho(\Phi)$. Therefore,
\begin{equation}
\label{corr2}
c_s^2=p'(\rho)=\frac{p'(\Phi)}{\rho'(\Phi)}=-\frac{\rho(\Phi)}{\int F'(\epsilon)\, d{\bf v}}=-\frac{\rho(\Phi)}{\int \frac{\partial f}{\partial v}(v,\Phi)\frac{1}{v}\, dv}.
\end{equation}
In the last expression, we have integrated over $v_1$...$v_{d-1}$ and noted $v$ for $v_d$, and $f(v,\Phi)$ for $\int f\, dv_1...dv_{d-1}$. For a spatially homogeneous system, the velocity of sound takes the simple form \cite{cvb,nyquist}:
\begin{equation}
\label{corr3}
c_s^2=-\frac{\rho}{\int_{-\infty}^{+\infty} \frac{f'(v)}{v}\, dv},
\end{equation}
where $f(v)$ is the reduced distribution of Section \ref{sec_vl}.

In general, the distribution function $f(v,\rho,T)$ and the pressure $p(\rho,T)$ depend on the density $\rho$ and on an external parameter $T$ which can be identified with a (generalized) temperature. Therefore, the velocity of sound is a function $c_s(\rho,T)$ of the density and of the temperature. Substituting this relation in the general equations (\ref{drv3}) and (\ref{asy7}), we can  express the stability criterion and the asymptotic distribution in terms of  $T$ and $\rho$. On the other hand, the total energy of a spatially  homogeneous system can be written as
\begin{equation}
\label{corr4}
E=\frac{1}{2}\int f v^2\, d{\bf r}d{\bf v}+W_0=\frac{d}{2}\int p\, d{\bf r}+W_0=\frac{d}{2}p(\rho,T)V+W_0,
\end{equation}
where $V$ is the volume and $W_0$ the potential energy in the homogeneous phase. From this equation, we can relate the energy to the temperature and express the results (\ref{drv3}) and (\ref{asy7}) in terms of $E$. Some explicit examples are given in Section \ref{sec_app}.

\section{Plasmas, self-gravitating systems, and the HMF model}
\label{sec_pgh}

In this Appendix, we provide explicit solutions of the dispersion relation for plasmas, self-gravitating systems, and for the HMF model. We consider the waterbag and the Cauchy distributions.

\subsection{Coulombian plasmas}

For a 3D Coulombian plasma, the potential of interaction is the solution of the Poisson equation $\Delta u=-4\pi(e^2/m^2)\delta({\bf x})$ yielding $(2\pi)^3\hat{u}(k)={4\pi e^2}/{m^2k^2}$. We introduce the plasma pulsation $\omega_P=({4\pi \rho e^2}/{m^2})^{1/2}$.

For the Cauchy distribution, the dispersion relation (\ref{c4}) becomes $\omega=\pm\omega_P-i u_0 k$. The perturbation oscillates with a pulsation $\omega_r=\pm\omega_P$ and is damped at a rate $\omega_i=-u_0 k<0$. For the long wavelengths, the damping is negligible and the  plasma oscillates with the proper pulsation $\omega_P$. This reflects the collective behavior of the system. The natural limit of collective behavior corresponds to $\omega_P\sim u_0 k$ leading to the Debye wavenumber $k_D=\omega_P/u_0=(4\pi\rho e^2/m^2 u_0^2)^{1/2}$.

For the waterbag distribution, the dispersion relation (\ref{w2}) becomes $\omega^2=v_m^2k^2+\omega_P^2$. The perturbation oscillates with a pulsation $\omega=\pm\sqrt{v_m^2k^2+\omega_P^2}$ without attenuation. For the long wavelengths, the plasma oscillates with the proper pulsation $\omega_P$. The Debye wavenumber is $k_D=\omega_P/v_m=(4\pi\rho e^2/m^2 v_m^2)^{1/2}$.

\subsection{Self-gravitating systems}

For a 3D self-gravitating system, the potential of interaction is the solution of the Poisson equation $\Delta u=4\pi G\delta({\bf x})$ yielding $(2\pi)^3\hat{u}(k)=-{4\pi G}/{k^2}$. We introduce the gravitational ``pulsation'' $\omega_G=\sqrt{4\pi G \rho}$.

For the Cauchy distribution, the dispersion relation (\ref{c4}) becomes $\omega=\pm i\omega_G-i u_0 k$. The Jeans wavenumber is $k_J=\omega_G/u_0=(4\pi G\rho/u_0^2)^{1/2}$. The system is stable for $k>k_J$ and unstable otherwise. In the stable case, the perturbation is damped with an exponential rate $\omega_i=\pm\omega_G-u_0 k<0$.
In the unstable case, the perturbation grows with an exponential rate $\omega_i=\omega_G-u_0 k>0$ (the other mode is damped with an exponential rate $\omega_i=-\omega_G-u_0 k<0$).

For the waterbag distribution, the dispersion relation (\ref{w2}) becomes $\omega^2=v_m^2k^2-\omega_G^2$. The Jeans wavenumber is $k_J=\omega_G/v_m=(4\pi G\rho/v_m^2)^{1/2}$.
The system is stable for $k>k_J$ and unstable otherwise. In the stable case, the perturbation oscillates with a pulsation $\omega_r=\pm\sqrt{v_m^2 k^2-\omega_G^2}$. In the unstable case, the perturbation grows with an exponential rate $\omega_i=\sqrt{\omega_G^2-v_m^2 k^2}>0$ (the other mode is damped with an exponential rate $\omega_i=-\sqrt{\omega_G^2-v_m^2 k^2}<0$).

\subsection{The repulsive HMF model}

For the repulsive HMF model, the potential of interaction is $u=\frac{1}{N}\left\lbrack 1+\cos(\theta-\theta')\right\rbrack$ yielding $\hat{u}_n=\frac{1}{2N}(2\delta_{n,0}+\delta_{n,1}+\delta_{n,-1})$. In a collisionless system (Vlasov), only the modes $n=\pm 1$ can propagate \cite{nyquist}.

For the Cauchy distribution, the dispersion relation (\ref{c4}) becomes $\omega=\pm {1}/{\sqrt{2}}-i u_0$. The perturbation oscillates  with a pulsation $\omega_r=\pm {1}/{\sqrt{2}}$ and is damped at a rate $\omega_i=-u_0<0$. For $u_0\rightarrow 0$, the damping is negligible and the system oscillates with the proper pulsation $1/\sqrt{2}$. This reflects the collective behavior of the system. The natural limit of collective behavior corresponds to $u_0^2=1/2$.

For the waterbag distribution, the dispersion relation  (\ref{w2}) becomes $\omega^2=v_m^2+1/2$. The perturbation oscillates with a pulsation $\omega=\pm\sqrt{v_m^2+1/2}$ without attenuation. For $v_m\rightarrow 0$, the system oscillates with the proper pulsation $1/\sqrt{2}$. The natural limit of collective behavior corresponds to $v_m^2=1/2$.

\subsection{The attractive HMF model}

For the attractive HMF model, the potential of interaction is  $u=\frac{1}{N}\left\lbrack 1-\cos(\theta-\theta')\right\rbrack$ yielding $\hat{u}_n=\frac{1}{2N}(2\delta_{n,0}-\delta_{n,1}-\delta_{n,-1})$. In a collisionless system (Vlasov), only the modes $n=\pm 1$ can propagate \cite{nyquist}.

For the Cauchy distribution, the dispersion relation (\ref{c4}) becomes $\omega=\pm i/\sqrt{2}-i u_0$. The system is stable for $u_0^2>1/2$ and unstable otherwise. In the stable case, the perturbation is damped with an exponential rate $\omega_i=\pm 1/\sqrt{2}-u_0<0$. In the unstable case, the perturbation grows with an exponential rate $\omega_i=1/\sqrt{2}-u_0>0$ (the other mode is damped with an exponential rate $\omega_i=-1/\sqrt{2}-u_0<0$).

For the waterbag distribution, the dispersion relation  (\ref{w2}) becomes $\omega^2=v_m^2-1/2$.
The system is stable for $v_m>1/\sqrt{2}$ and unstable otherwise. In the stable case, the perturbation oscillates with a pulsation $\omega_r=\pm\sqrt{v_m^2-1/2}$. In the unstable case, the perturbation grows with an exponential rate $\omega_i=\sqrt{1/2-v_m^2}>0$ (the other mode is damped with an exponential rate $\omega_i=-\sqrt{1/2-v_m^2}<0$).

\section{An identity}
\label{sec_identity}

For a collisionless system described by the Vlasov equation, the
Fourier-Laplace transform of the correlation function of the
fluctuations of the potential is given by Eq. (28) of
Ref. \cite{epjp}. Taking the inverse Laplace transform of this
equation and using Eq. (\ref{vl4}), we obtain
\begin{equation}
\label{identity1}
\langle \delta\hat{\rho}({\bf k},t)\delta\hat{\rho}({\bf k}',t')\rangle=\frac{m}{(2\pi)^d}\delta({\bf k}+{\bf k}')\int e^{-i{\bf k}\cdot {\bf v}(t-t')}\frac{f({\bf v})}{|\epsilon({\bf k},{\bf k}\cdot {\bf v})|^2}\, d{\bf v}.
\end{equation} 
The equilibrium correlation function is therefore
\begin{equation}
\label{identity2}
\langle \delta\hat{\rho}({\bf k})\delta\hat{\rho}({\bf k}')\rangle=\frac{m}{(2\pi)^d}\delta({\bf k}+{\bf k}')\int \frac{f({\bf v})}{|\epsilon({\bf k},{\bf k}\cdot {\bf v})|^2}\, d{\bf v},
\end{equation} 
where $f({\bf v})$ is the Maxwell distribution (\ref{app1}). On the other hand, the correlation function can be written as (see, e.g., Appendix A of \cite{jeanschemo}):
\begin{equation}
\label{identity3}
\langle \delta\hat{\rho}({\bf k})\delta\hat{\rho}({\bf k}')\rangle=\frac{\rho m}{(2\pi)^d}\left\lbrack 1+(2\pi)^d n \hat{h}({\bf k})\right\rbrack \delta({\bf k}+{\bf k}'),
\end{equation} 
where $\hat{h}({\bf k})$ is given by Eq. (\ref{asy9bb}). Therefore, we obtain the identity
\begin{equation}
\label{identity4}
\int \frac{f({\bf v})}{|\epsilon({\bf k},{\bf k}\cdot {\bf v})|^2}\, d{\bf v}=\rho\left\lbrack 1+(2\pi)^d n \hat{h}({\bf k})\right\rbrack=\frac{\rho}{\epsilon({\bf k},0)}=\frac{\rho}{1+(2\pi)^d\hat{u}(k)\beta\rho}.
\end{equation}

%\begin{acknowledgments}
%\end{acknowledgments}

% For figures use the graphicx package (distributed with LaTeX2e)
% and the \includegraphics macro defined in this package.
% For more details see e.g. the LaTeX Graphics Companion by Michel Goosens,
% Sebastian Rahtz, and Frank Mittelbach

% Here is an example of the general form of a figure:
% \begin{figure}
% \includegraphics{filename}%
% \caption{\label{}}
% \end{figure}
% Fill in the caption in the braces of the \caption{} command. Put the label
% that you will use with \ref{} command in the braces of the \label{} command.
% Use the figure* environment if the figure should span across the
% entire page instead of one column. There is no need to do explicit centering.

% If you have appendix consisting of several sections, uncomment
% the following command. Use \appendix* if there is only one section
% in the appendix
%\appendix
%\section{}

% Uncomment the following line if you use BibTeX to produce
% the reference section
% \bibliography{EE250sample}

\end{document}